\documentclass[aps,prd,twocolumn,groupedaddress,nofootinbib,longbibliography,floatfix]{revtex4-2}
\usepackage[utf8]{inputenc}

\usepackage{amsmath,amssymb}
\usepackage{graphicx}
\usepackage{siunitx}
\usepackage{dcolumn}
\usepackage{bm,color}
\usepackage{hyperref}
\usepackage{accents}
\usepackage{placeins}
\hypersetup{
    colorlinks = true,
    allcolors = blue
}
\usepackage{amssymb,float}
\usepackage{amsmath}
\usepackage{multirow}
\usepackage[parfill]{parskip}

\newcolumntype{K}[1]{>{\centering\arraybackslash}p{#1}}

\let\thempfootnote\thefootnote

\newcolumntype{A}{ >{$\quad} r <{$} @{} >{${}} l <{$} }
\makeatletter
\g@addto@macro\bfseries{\boldmath}
\makeatother

\begin{document}

\title{Spatial Curvature in $f(R)$ Gravity}


\author{Christine R. Farrugia}
\email[]{christine.r.farrugia@um.edu.mt}
\affiliation{Department of Mathematics, Faculty of Science, University of Malta, Msida MSD 2080, Malta}
\author{Joseph Sultana}
\email[]{joseph.sultana@um.edu.mt}
\affiliation{Department of Mathematics, Faculty of Science, University of Malta, Msida MSD 2080, Malta}
\author{Jurgen Mifsud}
\email[]{jurgen.mifsud@um.edu.mt}
\affiliation{Korea Astronomy and Space Science Institute, 776 Daedeokdae--ro, Yuseong--gu, Daejeon 34055, Republic of Korea\\
Institute of Space Sciences and Astronomy, University of Malta, Msida, MSD 2080, Malta}

\date{\today}

\begin{abstract}
In this work, we consider four $f(R)$ gravity models -- the Hu-Sawicki, Starobinsky, Exponential and Tsujikawa models -- and use a range of cosmological data, together with Markov Chain Monte Carlo sampling techniques, to constrain the associated model parameters. Our main aim is to compare the results we get when $\Omega_{k,0}$ is treated as a free parameter with their counterparts in a spatially flat scenario. The bounds we obtain for $\Omega_{k,0}$ in the former case are compatible with a flat geometry. It appears, however, that a higher value of the Hubble constant $H_0$ allows for more curvature. Indeed, upon including in our analysis a Gaussian likelihood constructed from the local measurement of $H_0$, we find that the results favor an open universe at a little over $1\sigma$. This is perhaps not statistically significant, but it underlines the important implications of the Hubble tension for the assumptions commonly made about spatial curvature. We note that the late-time deviation of the Hubble parameter from its $\Lambda$CDM equivalent is comparable across all four models, especially in the non-flat case. When $\Omega_{k,0}=0$, the Hu-Sawicki model admits a smaller mean value for $\Omega_{\text{cdm},0}h^2$, which increases the said deviation at redshifts higher than unity. We also study the effect of a change in scale by evaluating the growth rate at two different wavenumbers $k_\dagger$. Any changes are, on the whole, negligible, although a smaller $k_\dagger$ does result in a slightly larger average value for the deviation parameter $b$.

\end{abstract}


\maketitle


\section{Introduction}
\label{sec:intro}

Fourth-order metric theories of gravitation can be said to have originated from Weyl's 1918 non-integrable relativity theory \cite{Weyl}. This theory (or variants of it) was further investigated by scientists such as W.~Pauli, R.~Weitzenb\"{o}ck and F.~J\"{u}ttner, and served to introduce or promulgate key concepts such as conformal invariance, gravitational theories based on a geometrical approach, and the unification of the forces of Nature \cite{Schmidt2007}\footnote{Also refer to works cited therein.}. However, the popularity of Weyl's theory soon declined, namely due to the ambiguity of the associated Lagrangian and the problems posed by the higher order of the field equations. Additionally, there did not seem to be any experimental evidence against General Relativity (GR) that would favor the introduction of a more complicated theory. It was not until the 1970s that interest was revived. This happened as a result of factors such as the one-loop renormalizability of fourth-order metric theories, and the natural way in which inflation can be incorporated into them \cite{Schmidt2007}. Moreover, given a classical gravitational field arising from the energy-momentum tensor ($T_{\mu\nu}$) of quantized matter/radiation, the Lagrangian of fourth-order theories helps to erase any singularities that the gravitational interaction induces in $T_{\mu\nu}$ \cite{Utiyama}. 

The general class of \emph{Fourth-Order Gravity} is governed by an action whose gravitational part reads \cite{Capozziello2010, Stabile}
\begin{equation}
\mathcal{S}=\int \frac{\sqrt{-g}}{16\pi G}~f\left(R,R_{\alpha\beta}R^{\alpha\beta},R_{\alpha\beta\gamma\delta}R^{\alpha\beta\gamma\delta}\right)\text{d}^4x~,
\end{equation}
where $g$ is the determinant of the metric tensor $g_{\mu\nu}$, $R$, $R_{\alpha\beta}$ and $R_{\alpha\beta\gamma\delta}$ stand for the Ricci scalar and the Ricci and Riemann tensors, respectively, $f$ represents a generic analytical function, and $G$ is Newton's constant of gravitation. In the metric approach, the field equations are derived by varying the action with respect to $g^{\mu\nu}$ \cite{Stabile}. Among the theories obtained in this way is Conformal Weyl Gravity \cite{Mannheim1989}.

Another popular example is metric $f(R)$ gravity. Formulated by replacing the Ricci scalar in the GR action with a function thereof, $f(R)$ theory can be seen as a natural extension of GR \cite{Sultana2019}.  Despite its simplicity, however, it incorporates some of the basic characteristics of higher-order theories of gravity (i.e.~theories constructed from actions in which the $R$ of GR has been generalized to some function of higher-order curvature invariants), and is furthermore advantageous in that it appears to be the only higher-order theory that does not suffer from the Ostrogradski instability \cite{Sotiriou}. The prototype of $f(R)$ gravity has $f(R)=R-\alpha^4/R$ (where $\alpha\sim H_0$ and $H_0$ is the Hubble constant). It was adopted in an attempt to explain late-time cosmic acceleration \cite{Faraoni2008, Capozziello, SCarroll2004}, but has been ruled out on the basis of the Dolgov-Kawasaki instability \cite{Faraoni2008, Dolgov} and the fact that it does not have a viable weak-field limit \cite{Faraoni2008, Chiba2003}. In general, the class of models with $f(R)=R+\alpha R^{-n}$ cannot give rise to an acceptable cosmological expansion history for any $n>0$ or $n<-1$ \cite{Amendola2007}. 

The function associated with Starobinsky's inflationary model [$f(R)=R+\alpha R^2$] \cite{Starobinsky1980} was also one of the first to be proposed. Since then, $f(R)$ gravity has been the subject of numerous studies. One of its apparent benefits is the ability of certain models to reproduce both the early period of inflation and the current acceleration \cite{Cognola, Nojiri}. That said, due to the stringent constraints that a candidate model must satisfy -- for example, it has to predict a matter-dominated cosmic era -- only a few are still considered valid \cite{Sultana2019}. These are best tested on cosmological scales. Indeed, it is here that deviations from GR show up, so measurements of observables such as those related to galaxy clustering, the cosmic microwave background (CMB) or weak lensing are examples of pertinent cosmological probes \cite{DeFelice2010}. 

Among the viable $f(R)$ models are the ones put forward by Hu and Sawicki \cite{Hu2007}, Starobinsky \cite{Starobinsky}, Tsujikawa \cite{STsujikawa2008} and Cognola et al.~\cite{Cognola}. The said models are the subject of numerous works in the literature (see, for instance, \cite{Nunes, Romero, Sultana2019, Martinelli, Arjona, Odintsov2017, Hu, Linder, Geng, Basilakos2013}), but almost always in the context of a spatially flat universe. In fact, the assumption that spatial curvature is negligible is made by the greater majority of works in the literature, with the results of missions such as \emph{Planck} \cite{Planck2018VI}, WMAP \cite{WMAP} and SDSS \cite{Sanchez} usually used as justification. The constraints placed by the respective studies on the geometry of the Universe are indeed compatible with spatial flatness. It should be remembered, however, that they are obtained in the context of a $\Lambda$CDM cosmology. And even with regards to the standard model, some issues remain: the \emph{Planck} temperature (TT), polarization (EE + lowE) and temperature-polarization cross-correlation (TE) power spectra, for instance, appear to favor a closed universe\footnote{Adding the lensing reconstruction reduces this to a little more than $1\sigma$ \cite{Planck2018VI}.} at over $2\sigma$ \cite{Planck2018VI}, but adding baryon acoustic oscillation (BAO) data or measurements of the full-shape galaxy power spectrum [FS-P$(k)$] makes the results perfectly consistent with a flat universe. The caveat is that the TT, TE, EE + lowE data turns out to be in significant tension with both BAO and FS-P$(k)$ measurements when a curved universe is assumed (see Ref.~\cite{Vagnozzi} and works cited therein; using cosmic chronometers has been proposed as a solution \cite{Vagnozzi2021}). 

It is well--known that standard cosmic inflation predicts a flat geometry. This is because the curvature density parameter $\Omega_k$ decreases exponentially during the inflationary epoch, but only grows as a power law afterwards \cite{Planck2015XIII}. However, models of inflation that give rise to open \cite{Bucher, Linde} or closed universes \cite{Linde2003, Ratra} are also possible, although they often require a degree of fine tuning \cite{Planck2015XIII}. It has been suggested that spatial curvature could have emerged during the evolution of the Universe, once the growth of large-scale structure entered the non-linear regime. This conclusion was reached on the basis of the Silent Universe approximation \cite{Bolejko2018}. The emergence of curvature might additionally hold the key to a resolution of the currently unresolved tension between CMB and distance-ladder estimates of the Hubble constant \cite{Bolejko}.  

In view of this, the practice of setting $\Omega_{k,0}$ (the present-day value of $\Omega_k$) to zero appears somewhat premature. There is also the fact that more stringent constraints on $\Omega_{k,0}$ could serve as important tests of eternal inflation models (see \cite{Leonard} and references therein). Furthermore, large-scale structure effects (such as those due to local inhomogeneities) could bias our measurements and shift the inferred $\Omega_{k,0}$ from the background value unless properly accounted for \cite{Bull, Leonard}, as could higher-order perturbations like second-order lensing corrections \cite{Leonard}. Another point to keep in mind is the strong degeneracy that frequently exists between dark energy parameters and $\Omega_{k,0}$. Many times, the problem is circumvented either by setting the latter to zero, or by only considering specific classes of the former. A case in point is the dark energy equation-of-state (EoS) parameter, $w_\text{de}$, for which a functional form is usually assumed. The result is that spatial curvature is mostly studied in a rather restrictive framework. Of particular concern is the fact that if the true value of $\Omega_{k,0}$ deviates from zero, assuming a flat geometry induces errors in $w_\text{de}$ that grow rapidly with redshift, even if the curvature is in reality only very small \cite{Clarkson}. 

In this work, we consider four $f(R)$ models and use observational data to place bounds on cosmological and model-specific parameters. Our main aim is to investigate how constraints are affected if $\Omega_{k,0}$ is treated as a free parameter. First we go over the preliminary theory (Section \ref{sec:preliminaries}), then introduce the relevant likelihoods in Section \ref{sec:observational}. Results are presented and discussed in Section \ref{sec:results}, while Section \ref{sec:conclusion} is dedicated to the concluding remarks. We use units in which the speed of light in vacuum, $c$, is equal to unity. 

\section{Metric $f(R)$ Gravity: Preliminaries}
\label{sec:preliminaries}
\subsection{The field equations}
\label{subsec:field_eqs}
At the basis of $f(R)$ theory is a generalization of the Einstein-Hilbert action of GR to:
\begin{equation}
\mathcal{S}=\int \frac{\sqrt{-g}}{16\pi G}\,f(R)\,\text{d}^4 x~.
\label{action}
\end{equation}
Here, $f(R)$ is a generic function of the Ricci curvature scalar. The field equations are obtained by varying the action with respect to the inverse metric tensor $g^{\mu\nu}$ \cite{Faraoni2008}, and collectively read
\begin{align}
\left(G_{\mu\nu}+\frac{1}{2}Rg_{\mu\nu}-\nabla_\mu\nabla_\nu+g_{\mu\nu}\Box\right)f_R -\frac{1}{2}f(R)g_{\mu\nu}&\notag\\[0.3em]=8\pi G\,T_{\mu\nu}&~,
\label{field_eq}
\end{align}
where $G_{\mu\nu}$ is the Einstein tensor ($G_{\mu\nu}=R_{\mu\nu}-Rg_{\mu\nu}/2$), $f_R=\text{d} f/\text{d} R$, the quantity $\nabla_\mu$ represents the covariant derivative operator constructed from the metric connection, and $\Box\equiv g_{\mu\nu}\nabla^{\mu}\nabla^{\nu}$. We shall model the matter/energy content of the Universe as a perfect fluid with proper density $\rho$, corresponding isotropic pressure $p$ and four-velocity $u^\mu$. The energy-momentum tensor of such a fluid reads
\begin{equation}
T_{\mu\nu}=(\rho+p)u_\mu u_\nu+p g_{\mu\nu}~.
\label{energy_momentum}
\end{equation}
In a Friedmann-Lema\^{i}tre-Robertson-Walker (FLRW) cosmology, the field equations [i.e.~Eq.~(\ref{field_eq})] can be recast into the form of their General Relativistic counterparts \cite{Nunes}. We may therefore write:
\begin{align}
H^2&=\frac{8\pi G}{3} \rho_\text{tot}-\frac{\kappa}{a^2}~,\label{Friedmann1}\\[0.3em]
\frac{\ddot{a}}{a}&=-\frac{4\pi G}{3}\left(\rho_\text{tot}+3p_\text{tot}\right)~.\label{Friedmann2}
\end{align}
In the above, $a$ is the scale factor, normalized with respect to its present-day value, and $H=\dot{a}/a$ is the Hubble parameter; an overdot denotes differentiation with respect to cosmic time $t$. The parameter $\kappa$ represents the spatial curvature and has dimensions of $\text{length}^{-2}$, while $\rho_\text{tot}=\rho+\rho_\text{de}$ and $p_\text{tot}=p+p_\text{de}$, $\rho$ and $p$ being the energy density and pressure from Eq.~(\ref{energy_momentum}). However, while in GR $\rho_\text{de}$ and $p_\text{de}$ are attributes of a physical component -- namely, the vacuum energy we denote by $\Lambda$ -- in $f(R)$ theory they may be expressed as a collection of terms which result from the modification to the geometry of the space-time manifold:
\begin{align}
&8\pi G\rho_\text{de}=\notag\\[0.3em]&\frac{1}{2}(f_R R-f)-3H\dot{f}_R+3(1-f_R)H^2+\frac{3\kappa}{a^2}(1-f_R)~,\label{rho_de}\\[0.3em]
&8\pi G p_\text{de}=\,\frac{1}{2}(f-f_R R)+\ddot{f}_R-(1-f_R)(2\dot{H}+3H^2)+\notag\\[0.3em]&2H\dot{f}_R+\frac{\kappa}{a^2}(f_R-1)~.
\end{align}
Simply put, the impact of these geometric terms on cosmic dynamics mimics the effects of a dark energy component with density $\rho_\text{de}$ and pressure $p_\text{de}$ \cite{Nunes}. We additionally note that this effective dark energy does not interact with matter or radiation. Consequently, conservation of energy implies that:
\begin{equation}
\dot{\rho}+3H(\rho+p)=0~,
\end{equation}
where $\rho=\rho_\text{m}$ (or $\rho_\text{r}$) and $p=p_\text{m}$ (or $p_\text{r}$). The subscripts `m' and `r' denote matter (cold dark matter and baryons) and radiation (photons and massless neutrinos), respectively. 

As mentioned earlier, a valid $f(R)$ theory must fulfill a number of criteria \cite{Sotiriou, Faraoni2008, Faraoni2006, Appleby}. Firstly, since the quantity $G_\text{eff}\equiv G/f_R$ acts as an effective gravitational coupling, the requirement that the graviton carries positive kinetic energy implies that $G_\text{eff}>0$, which in turn imposes the bound $f_R > 0$. Secondly, avoiding instabilities of the Dolgov-Kawasaki type \cite{Dolgov} necessitates that $\text{d}^2f/\text{d}R^2 \geq 0$. As for the cosmological dynamics, the theory should behave like $\Lambda$CDM at high redshifts, because the standard model is well-supported by CMB data in this regime. We therefore expect that $\lim_{R\to\infty}f(R)=R+\text{constant}$. A late-time expansion history similar to the one in a $\Lambda$CDM cosmology is also desirable, albeit in the absence of a cosmological constant; that is to say, viable $f(R)$ models should satisfy the condition $\lim_{R\to0}f(R)=R+0$ \cite{Hu2007}. 

The successes of $\Lambda$CDM on Solar-System scales suggest that its phenomenology should be a limiting case \cite{Hu2007} of any sound alternative theory. In metric $f(R)$, however, the Ricci curvature introduces a scalar degree of freedom, which could cause post-Newtonian constraints obtained from Solar System experiments to be violated. The model only remains valid if the scalar field can somehow be `shielded' from such experiments. This may be achieved via the so-called \emph{chameleon mechanism}, whereby the effective mass $M$ of the scalar varies according to the energy density of the local environment. In high-density regions like the Solar System, a large $M$ would shorten the range of the scalar field to scales that cannot currently be probed by weak-field experiments. On the other hand, $M$ would have to be small at cosmological densities, so as to allow the scalar field to act over a long range and drive the acceleration of the Universe \cite{Sotiriou, Khoury, Brax}. One important thing to note about chameleon behavior is that it cannot be described as a fine-tuning mechanism. Rather, it is a natural and intrinsic property of those $f(R)$ models whose weak-field limit satisfies observational constraints. 

Phase space analysis can also yield a wealth of information. In a particularly note-worthy study that takes this approach \cite{Amendola2007}, the authors consider the quantities $m=[Rf_{RR}/f_R](r)$  ($f_{RR}$ stands for $\text{d}^2f/\text{d}R^2$) and $r=-Rf_R/f$ and investigate the behavior of the $m(r)$ curve in the $(r,m)$ plane. It is found that for an $f(R)$ model to admit a viable matter-dominated epoch, the curve should satisfy the conditions $m(r)\approx +0$ and $\text{d}m/\text{d}r>-1$ at $r\approx -1$. Additionally, a valid period of late-time acceleration requires that $m=-r-1$ while $(\sqrt{3}-1)/2<m\leq 1$ and $\text{d}m/\text{d}r<-1$, or that $m$ lies in the range $(0,1]$ at $r=-2$ \cite{Amendola2007}.

\subsection{The cosmological equations as a set of first-order differential equations}
\label{subsec:1st_order_ODEs}
To avoid instabilities when solving Eqs.~(\ref{Friedmann1}) and (\ref{Friedmann2}) numerically, we rewrite them as a set of first order ordinary differential equations. To this end, we follow Refs.~\cite{Cruz-Dombriz, Carloni, Amendola2007, Abdelwahab, Odintsov}. The starting-point is the change of variables given by:
\begin{alignat}{2}
s&=\frac{R}{6\left(H_0^{\Lambda}\,\eta\right)^2}~;~~~~~\qquad x&&=-R'(z)(1+z)~;\notag\\[0.3em]
y&=\frac{f(R)}{6f_R\left(H_0^{\Lambda}\,\eta\right)^2}~;~~~~~~\omega_\text{m}&&=\frac{\Omega_{\text{m},0}^\Lambda(1+z)^3}{\eta^2f_R}~;\notag\\[0.3em]
\omega_\text{r}&=\frac{\Omega_{\text{r},0}^\Lambda(1+z)^4}{\eta^2f_R}~;~~~~~\quad K&&=\frac{\kappa(1+z)^2}{\left(H_0^\Lambda\,\eta\right)^2}~.
\label{change_of_var}
\end{alignat}
Here, a prime denotes differentiation with respect to the argument and $z$ is the cosmological redshift, while $\eta$ is defined as the ratio $H/H_0^{\Lambda}$. $\Omega_{\text{m},0}$ and $\Omega_{\text{r},0}$ are the values of the matter and radiation density parameters at $z=0$, respectively, and a superscript $\Lambda$ indicates quantities as measured/inferred in the framework of a $\Lambda$CDM cosmology. 

We have already seen that a candidate function $f(R)$ ideally satisfies the condition $\lim_{R\to\infty}f(R)=R+\text{constant}$. This implies that Eq.~(\ref{action}) becomes indistinguishable from the Einstein-Hilbert action at high redshifts, since the latter has Lagrangian density $(\sqrt{-g}/16\pi G)(R-2\Lambda)$. Consequently, at early times the $f(R)$ cosmology behaves as a $\Lambda$CDM model \emph{having cosmological constant $\varLambda=-\text{constant}/\mathit{2}$}. Let us reinterpret the superscript $\Lambda$ as a label for the parameters of this specific $\Lambda$CDM model. If we write the quantity $-\text{constant}/2$ as $\Lambda^{f(R)}$, it follows that $\Lambda^{\Lambda}=\Lambda^{f(R)}$, and hence that
\begin{equation}
\left(H_0^\Lambda\right)^2\Omega_{\Lambda,0}^\Lambda=\left(H_0^{f(R)}\right)^2\Omega_{\Lambda,0}^{f(R)}~,
\end{equation}
where $\Omega_{\Lambda,0}$ is the present-day value of the density parameter associated with $\Lambda$. Furthermore, given that the matter component is described by the same energy-momentum tensor in both $\Lambda$CDM and $f(R)$ gravity, and assuming that the two theories should lead to the same physical matter density today, we obtain the relation 
\begin{equation}
\Omega_{\text{m},0}^{\Lambda}\left(H_0^\Lambda\right)^2 = \Omega_{\text{m},0}^{f(R)}\left(H_0^{f(R)}\right)^2 = \frac{8\pi G}{3}\rho_\text{m}(z=0)~.
\label{OmLCDMfR}
\end{equation}
In general, though, 
\begin{equation}
H_0^{f(R)}\neq H_0^\Lambda\quad\text{and}\quad\Omega_{\text{m},0}^{f(R)}\neq\Omega_{\text{m},0}^\Lambda~,
\end{equation}
since the two models are expected to diverge at late times \cite{Odintsov2017,Hu2007}.

A similar conclusion can be reached for the radiation density:
\begin{equation}
\Omega_{\text{r},0}^{\Lambda}\left(H_0^\Lambda\right)^2 = \Omega_{\text{r},0}^{f(R)}\left(H_0^{f(R)}\right)^2~;\qquad \Omega_{\text{r},0}^{\Lambda} \neq \Omega_{\text{r},0}^{f(R)}~.\label{OrLCDMfR}
\end{equation}
In order to account for spatial curvature, it is customary to introduce a quantity $\Omega_{k,0}$ that complements $\Omega_\text{m,0}$ and $\Omega_{\text{r},0}$ and is equal to $-\kappa/H_0^2$. The curvature parameter $\kappa$ is a constant, so at early times an $f(R)$ model mimics a $\Lambda$CDM cosmology having $\kappa^\Lambda=\kappa^{f(R)}$. Therefore, it follows that
\begin{equation}
\Omega_{k,0}^{\Lambda}\left(H_0^\Lambda\right)^2 = \Omega_{k,0}^{f(R)}\left(H_0^{f(R)}\right)^2~.\label{OkLCDMfR}
\end{equation}
$\kappa$ is defined as the ratio $k/\mathcal{R}_0^2$, $k$ being the normalized curvature parameter (equal to $\pm 1$ or 0) and $\mathcal{R}_0$ the present-day value of the non-normalized scale factor. So if $\kappa^\Lambda=\kappa^{f(R)}$, it must mean that at high redshifts, an $f(R)$ model with current scale factor $\hat{\mathcal{R}}_0$ behaves as a $\Lambda$CDM model that also has $\mathcal{R}_0=\hat{\mathcal{R}}_0$.

Eqs.~(\ref{OmLCDMfR})--(\ref{OkLCDMfR}) make it possible to rewrite $H_0^{f(R)}$, $\Omega_{\text{m},0}^{f(R)}$, $\Omega_{\text{r},0}^{f(R)}$ and $\Omega_{k,0}^{f(R)}$ in terms of their $\Lambda$CDM counterparts [as was already done for the expressions in  Eq.~(\ref{change_of_var})]. This is especially convenient, because it enables us to construct informative priors for the $f(R)$ cosmological parameters using \emph{Planck} constraints \cite{Planck2018VI}, which makes the process of sampling the parameter space much more efficient.  

Let us now return to Eq.~(\ref{change_of_var}). In terms of the new variables ($\eta$, $s$, $x$, $y$, $\omega_\text{m}$, $\omega_\text{r}$ and $K$), the system of cosmological equations to be solved becomes:
\begin{align}
\eta'(z)=\,&\frac{\eta}{z+1}(2-s+K)~;\label{etaprime}\\[0.3em]
s'(z)=\,&-\frac{s}{z+1}\left(\frac{x}{R}+4-2s+2K\right)~;\\[0.3em]
x'(z)=\,&\frac{1}{\Gamma(z+1)}\left[(x\Gamma)^2+s(x\Gamma-1)+3y-1+\omega_\text{r}-\right.\notag\\[0.3em]&\left.K(1+x\Gamma)\right]-x\Gamma'(z)\Gamma^{-1}~;\label{xprime}\\[0.3em]
y'(z)=\,&-\frac{1}{z+1}\left[s\frac{x}{R}+y(4-x\Gamma-2s+2K)\right]~;\\[0.3em]
\omega_\text{m}'(z)=\,&\frac{\omega_\text{m}}{z+1}(x\Gamma+2s-2K-1)~;\label{ommprime}\\[0.3em]
\omega_\text{r}'(z)=\,&\frac{\omega_\text{r}}{z+1}(x\Gamma+2s-2K)~;\label{omega_r}\\[0.3em]
K'(z)=\,&-\frac{2K}{z+1}(K-s+1)~,\label{K}
\end{align}
where $\Gamma$ is equal to $f_{RR}/f_R$ and serves to identify the particular $f(R)$ model.

\subsection{Specific $f(R)$ models}\label{subsec:fR_models}
Among the viable $f(R)$ models are the ones put forward by Hu and Sawicki \cite{Hu2007}, Starobinsky \cite{Starobinsky}, Tsujikawa \cite{STsujikawa2008} and Cognola et al.~\cite{Cognola}. In this sub-section, we take a closer look at each of them in turn.

\subsubsection{The Hu-Sawicki model}
Hu and Sawicki proposed a class of `broken power-law' models \cite{Hu2007}:
\begin{equation}
f(R)_\text{HS}^{}=R-\mu^2\frac{c_1\left(R/\mu^2\right)^{n_\text{HS}^{}}}{1+c_2\left(R/\mu^2\right)^{n_\text{HS}^{}}}~.
\label{HS}
\end{equation}
Here, $c_1$ and $c_2$ are dimensionless parameters, $n_\text{HS}^{}$ represents a positive constant that is usually assumed to be an integer, and $\mu^2\approx \Omega_{\text{m},0}^{}\,H_0^2$.

It may be shown that the Hu-Sawicki model includes $\Lambda$CDM as a limiting case and can, in fact, be seen as a late-time modification of the latter \cite{Romero}. Moreover, it is possible to explicitly incorporate the cosmological constant $\Lambda$ into Eq.~(\ref{HS}) by making the substitutions \cite{Basilakos2013}
\begin{equation}
\Lambda=\frac{\mu^2c_1}{2c_2}~;~~~~b=\frac{2c_2^{1-n_\text{HS}^{-1}}}{c_1}~,
\end{equation}
which cast $f(R)_\text{HS}^{}$ into the form \cite{Basilakos2013, Bamba2013}:
\begin{equation}
f(R)_\text{HS}^{}=R-2\Lambda\left(1-\frac{1}{1+[R/(b\Lambda)]^{n_\text{HS}^{}}}\right)~.
\label{HS2}
\end{equation}
Eq.~(\ref{HS2}) makes it apparent that at high redshifts, when $R\gg\Lambda$, $f(R)_\text{HS}^{}$ reduces to $R-2\Lambda$ and $\Lambda$CDM is consequently recovered \cite{Bamba2013}. The differences that emerge at lower redshifts are quantified by the \emph{deviation parameter} $b$ \cite{Basilakos2013} ($b=0$ corresponds to $\Lambda$CDM). Constraints placed on $b$ by means of cosmological data, therefore, translate into bounds on the allowed variation from the standard model. Additionally, the time at which these variations set in is controlled by $n_\text{HS}^{}$: the larger the value of this parameter, the longer it takes for the Hu-Sawicki model to diverge from $\Lambda$CDM \cite{Hu2007}. We shall follow other works in the literature and (without loss of generality) set $n_\text{HS}^{}$ to unity \cite{Nunes, Romero, Basilakos2013}. Furthermore, only non-negative values of $b$ will be considered. The reason is that when $n_\text{HS}^{}=1$, $f_{RR}=4b\Lambda^2/(R+b\Lambda)^3$, and so having $b<0$ would mean that $f_{RR}$ becomes negative as soon as $R>-b\Lambda$. We have already seen that viable $f(R)$ models have $f_{RR}\geq 0$.

Before proceeding to the next model, it would be interesting -- and extremely useful for setting up the numerical procedures performed later -- to determine at what redshift the Hu-Sawicki model becomes effectively indistinguishable from $\Lambda$CDM. To this end, we adopt a procedure similar to the one proposed in Ref.~\cite{Odintsov2017} for exponential $f(R)$. Eq.~(\ref{HS2}) allows us to deduce that if $f(R)_\text{HS}^{}$ is to approach $R-2\Lambda$ at high redshifts, the magnitude of $1/\{1+[R^{f(R)}/(b\Lambda)]^{n_\text{HS}^{}}\}$ must decrease asymptotically to zero as we go back in time. This may be expressed as the requirement that at some redshift $z_\text{bound}$, the quantity $1/(1+[R^{f(R)}/(b\Lambda)]^{n_\text{HS}^{}})$ is equal to $\epsilon$ (with $\epsilon\ll 1$), at which point any differences between the $f(R)$ model and $\Lambda$CDM are negligible. Moreover, at $z=z_\text{bound}$ one expects $R^{f(R)}$ to take the form $R^\Lambda+\xi$ (for some $|\xi| \ll R^\Lambda$). With these considerations in mind, we may write
\begin{align}
&1+\left(\frac{R^{\Lambda}+\xi}{b\Lambda}\right)^{n_\text{HS}^{}}\approx\,1+\left(\frac{R^{f(R)}}{b\Lambda}\right)^{n_\text{HS}^{}}=\,\frac{1}{\epsilon}~;\notag\\[0.5em]
\implies &\left(\frac{b\Lambda}{R^\Lambda+\xi}\right)^{n_\text{HS}^{}}\approx\,\frac{\epsilon}{1-\epsilon}~;\notag\\[0.5em]
\implies &\,\,\frac{b\Lambda}{R^\Lambda+\xi}\approx\,\left(\frac{\epsilon}{1-\epsilon}\right)^{n_\text{HS}^{-1}}=\nu~;\notag\\[0.5em]
\implies &\,\,\,\frac{b\Lambda}{R^\Lambda}\approx\,\nu\left(1+\frac{\xi}{R^\Lambda}\right)~,\notag
\end{align}
and if terms higher than first order in $\epsilon$ or $\xi/R^\Lambda$ are discarded, it follows that
\begin{align}
\nu&=\left(\frac{\epsilon}{1-\epsilon}\right)^{n_\text{HS}^{-1}} = \epsilon^{n_\text{HS}^{-1}}(1-\epsilon)^{-n_\text{HS}^{-1}}\notag\\[0.3em]&\approx \epsilon^{n_\text{HS}^{-1}}\left(1+\frac{\epsilon}{n_\text{HS}}\right)\approx \epsilon^{n_\text{HS}^{-1}}~;\notag\\[0.3em]
\frac{b\Lambda}{R^\Lambda}&\approx\,\nu\left(1+\frac{\xi}{R^\Lambda}\right)\approx\epsilon^{n_\text{HS}^{-1}}\left(1+\frac{\xi}{R^\Lambda}\right)\approx\epsilon^{n_\text{HS}^{-1}}\approx\nu~.
\end{align}
Therefore, $b\Lambda/R^\Lambda\approx\,\nu$ at $z=z_\text{bound}$. Using the relation $R^\Lambda=3H_0^2[\Omega_{\text{m},0}(1+z)^3+4\Omega_{\Lambda,0}]$, we solve for $z_\text{bound}$ and find that
\begin{equation}
z_\text{bound}=\left[\frac{\Omega_{\Lambda,0}}{\Omega_{\text{m},0}}\left(\frac{b}{\nu}-4\right)\right]^{1/3}-1~.
\label{z_bound}
\end{equation}
In the case of the Hu-Sawicki model, the ratio $b\Lambda/R^\Lambda$ (henceforth referred to as $\nu$) equates to $[\epsilon/(1-\epsilon)]^{n_\text{HS}^{-1}}$ at $z=z_\text{bound}$. The remaining models will give rise to different expressions for $b\Lambda/R^\Lambda$, which will all be functions of the quantity $\epsilon$ obtained by putting $f(R[z=z_\text{bound}])=R-2\Lambda(1-\epsilon)$ (thus, $\epsilon$ differs from model to model). All we need to remember, however, is that $\epsilon$ is a positive constant much smaller than unity. More details about the values we choose for $\epsilon$ are given in Section \ref{sec:observational}.

\subsubsection{The Starobinsky model}
Starobinsky proposed the function \cite{Starobinsky}:
\begin{equation}
f(R)_\text{S}^{}=R+\lambda R_\text{S}^{}\left[\left(1+\frac{R^2}{R_\text{S}^2}\right)^{-n_\text{S}^{}}-1\right]~,
\end{equation}
where $n_\text{S}^{}$ and $\lambda$ denote positive constants, and the third constant, $R_\text{S}^{}$, is expected to be of the order of the present-day Ricci scalar \cite{STsujikawa2008}. We write $f(R)_\text{S}^{}$ in the form of a perturbed $\Lambda$CDM Lagrangian \cite{Nunes}:
\begin{equation}
f(R)_\text{S}^{}=R-2\Lambda\left[1-\left(1+\frac{R^2}{(b\Lambda)^2}\right)^{-n_{\text{S}^{}}}\right]~,
\end{equation}
which clearly shows that $f(R)_\text{S}^{}\rightarrow R-2\Lambda$ when $R\gg\Lambda$ or when $b\rightarrow0$. $\Lambda$ and $b$ may be expressed in terms of the original parameters as $\lambda R_\text{S}^{}/2$ and $2/\lambda$, respectively \cite{Nunes}.

The redshift $z_\text{bound}$ is again given by Eq.~(\ref{z_bound}). Now, however, we have that
\begin{equation}
\nu=\sqrt{\frac{\epsilon^{1/n_\text{S}^{}}}{1-\epsilon^{1/n_\text{S}^{}}}}~~.
\end{equation}

Without loss of generality, we shall put $n_\text{S}$ equal to unity from now on.

\subsubsection{The Exponential model}
In this case, the $f(R)$ function reads \cite{Cognola}
\begin{equation}
f(R)_\text{E}^{}=R+\beta\left[\text{exp}(-\gamma R)-1\right]~,
\end{equation}
or equivalently \cite{Nunes}
\begin{equation}
f(R)_\text{E}^{}=R-2\Lambda\left[1-\text{exp}\left(-\frac{R}{b\Lambda}\right)\right]~,
\end{equation}
with $\Lambda=\beta/2$ and $b=2/(\gamma\beta)$. $\beta$ and $\gamma$ are two constants that characterize the model; $\gamma$ must be positive so that $b\geq 0$ and at high redshifts, when $R\gg\Lambda$, the exponential function becomes negligible and $\Lambda$CDM is recovered \cite{Odintsov2017}. This also happens as $b\rightarrow0$.

The redshift $z_\text{bound}$ may be estimated from Eq.~(\ref{z_bound}) by making use of the relation
\begin{equation}
\nu=\frac{1}{\ln{(1/\epsilon)}}~~.
\end{equation}

\subsubsection{The Tsujikawa model}
The model proposed by Tsujikawa is based on the function \cite{STsujikawa2008}
\begin{equation}
f(R)_\text{T}^{}=R-\zeta R_\text{T}^{}\tanh{\left(\frac{R}{R_\text{T}^{}}\right)}~,
\end{equation}
where $\zeta$ and $R_\text{T}$ are positive constants. We may alternatively write
\begin{equation}
f(R)_\text{T}^{}=R-2\Lambda\tanh{\left(\frac{R}{b\Lambda}\right)}~.
\end{equation}
Here, $b=2/\zeta$ and $\Lambda=\zeta R_\text{T}^{}/2$ \cite{Nunes}, and the model becomes equivalent to $\Lambda$CDM either when $R\gg\Lambda$ or when $b\rightarrow0$ (since $\tanh{[R/(b\Lambda)]}\rightarrow 1$ in both cases). The quantity $\nu$ required to calculate $z_\text{bound}$ [Eq.~(\ref{z_bound})] is given by
\begin{equation}
\nu=\frac{1}{\text{arctanh}(1-\epsilon)}~~.
\end{equation}

\subsection{Perturbations in $f(R)$ Gravity}
We start by considering the perturbed field equations:
\begin{align}
&\delta G^\mu_\nu f_R + \left(R^\mu_\nu-\nabla^\mu\nabla_\nu+\delta^\mu_\nu\Box\right)f_{RR}\,\delta R + (\delta g^{\mu\alpha}\nabla_\nu\nabla_\alpha -\notag\\[0.3em]&\delta^\mu_\nu \delta g^{\alpha\beta}\nabla_\alpha\nabla_\beta)f_R\,+\left(g^{\alpha\mu}\delta\Gamma^\gamma_{\alpha\nu}-\delta^\mu_\nu g^{\alpha\beta}\delta\Gamma^\gamma_{\beta\alpha}\right)\partial_\gamma f_R\notag\\[0.3em]&=8\pi G\,\delta T^\mu_\nu~.
\label{field_eq_pert_fR}
\end{align}
In the above, the quantities $\delta G^\mu_\nu$, $\delta R$, $\delta g_{\mu\nu}$ and $\delta\Gamma^\mu_{\sigma\nu}$ denote perturbations in the Einstein tensor, the Ricci scalar, the metric tensor and the metric connection, respectively, while $\delta^\mu_\nu$ is the Kronecker delta. Perturbations in quantities related to the geometry of the space-time manifold appear on the left-hand side of Eq.~(\ref{field_eq_pert_fR}). Meanwhile, the right-hand side constitutes the perturbed part $\delta T^\mu_\nu$ of the energy-momentum tensor, and may be expanded as follows \cite{Bardeen, Bertschinger}:
\begin{align}
\delta T^0_0&=-\delta \rho_\text{m}~;\qquad\delta T^0_i=(\rho_\text{m}+p_\text{m})v_i~;\notag\\[0.3em]\delta T^i_0&=-(\rho_\text{m}+p_\text{m})v^i~;\qquad\delta T^i_j=\delta p_\text{m}\, \delta^i_j~.
\label{energy-momentum}
\end{align}
Here, $\rho_\text{m}$ and $p_\text{m}$ are the background values of the matter energy density and pressure, respectively, and $\delta\rho_\text{m},\,\delta p_\text{m}$ their associated perturbations. The 3-vector $v^i$ represents the perturbation in the spatial velocity. We keep to the perfect-fluid form and hence do not consider anisotropic stresses (which explains why $\delta T^i_j=0$ for $i\neq j$) \cite{Mukhanov}.

In the conformal Newtonian gauge, the perturbed FLRW metric takes the form:
\begin{equation}
\text{d}s^2=a^2(\tau)\left[-(1+2\Phi)\text{d}\tau^2+\gamma_{ij}(1-2\Psi)\text{d}x^i\text{d}x^j\right]~,
\label{metric-pert}
\end{equation}
where $\gamma_{ij}=\delta_{ij}\left[1+\frac{1}{4}\kappa\left(x^2+y^2+z^2\right)\right]^{-2}$ \cite{Mukhanov}, $\tau$ is the conformal time (which is related to the cosmic time $t$ via the scale factor: $\text{d}\tau=\text{d}t/a$), and we have made use of quasi-Cartesian coordinates \cite{Weinberg}.\footnote{In 3D Euclidean space with a Cartesian coordinate system, the 3-metric $\gamma_{ij}$ has components $\delta_{ij}$, rather than $\delta_{ij}\left[1+\frac{1}{4}\kappa\left(x^2+y^2+z^2\right)\right]^{-2}$. We shall be using $\vec{x}$ as shorthand for the spatial vector $(x,y,z)$.} The scalar potentials $\Phi(\tau, \vec{x})$ and $\Psi(\tau, \vec{x})$ constitute the metric perturbations in our particular gauge; they are assumed to satisfy the condition $|\Phi|,\,|\Psi|\ll1$.

The next step involves modeling perturbations as wave functions in momentum space. A perturbation $\delta g(\tau,\vec{x})$ in physical space translates into a sum (or integral) over $k_\dagger$-modes in momentum space. For instance, in the absence of spatial curvature, we have that 
\begin{equation}
\delta g(\tau,\vec{x})=\sum_{k_\dagger} \delta \hat{g}(\tau,k_\dagger)\,\text{e}^{i\vec{k}_\dagger\boldsymbol{\cdot}\vec{x}}~.
\label{Fourier}
\end{equation}
Each of the modes in question has a characteristic comoving wave vector $\vec{k}_\dagger$ (and corresponding wave number $k_\dagger=|\vec{k}_\dagger|$). Since we consider perturbations to linear order only, it follows that if $\delta g(\tau,\vec{x})$ satisfies a particular equation, then the individual modes summed over in Eq.~(\ref{Fourier}) also satisfy that equation, albeit for different values of $k_\dagger$. In other words, perturbations with a different wave number decouple, and so we can write our equations in terms of a generic mode $\delta \hat{g}(\tau,k_\dagger)Q(\vec{x},k_\dagger)$ \cite{Abbott1986, Bardeen}. In spherical coordinates, the purely spatial part of each mode of oscillation is given by
\begin{equation}
    Q(\vec{x},k_\dagger) = \Theta^\ell_{\beta(k_\dagger)}(r)Y_{\ell m}(\theta,\phi)~,
\label{Q}    
\end{equation}
where $Y_{\ell m}$ denotes the spherical harmonics and the form of the function $\Theta^\ell_\beta$ depends on the value of $\kappa$ (refer to \cite{Abbott1986} and \cite{Harrison} for more details). It may be shown that in the flat case, the right-hand side of Eq.~(\ref{Q}) reduces to $\text{e}^{i\vec{k}_\dagger\boldsymbol{\cdot}\vec{x}}$.

In momentum space, then, the time-time component of Eq.~(\ref{field_eq_pert_fR}) reads \cite{Hwang}:
\begin{align}
&2f_R\left\{\Psi\left(k_\dagger^2-3\kappa\right)+3\mathcal{H}\left[\Psi'(\tau)+\Phi\mathcal{H}\right]\right\}+f_{RR}[3\mathcal{H}'(\tau)\delta R\notag\\[0.3em]&-k_\dagger^2\delta R - 3\mathcal{H}\,\delta R'(\tau)]-3\mathcal{H}\,\delta{R}\,f_{RR}'(\tau)+3f_R'(\tau)[2\mathcal{H}\Phi\notag\\[0.3em]&+\Psi'(\tau)]+8\pi G a^2\rho_\text{m}\delta_\text{m}=0~,
\label{perturb_5}
\end{align}
with
\begin{align}
\delta R&=\frac{2}{a^2}\bigg\{k_\dagger^2(\Phi-2\Psi)-3[2\Phi\mathcal{H}'(\tau)+3\mathcal{H}\Psi'(\tau)+\mathcal{H}\Phi'(\tau)\notag\\[0.3em]&-2\kappa \Psi + \Psi''(\tau)+2\Phi\mathcal{H}^2]\bigg\}~,
\label{deltaR}
\end{align}
where $\mathcal{H}$ is the conformal Hubble parameter [equivalent to the ratio $a'(\tau)/a$], and the matter density contrast function, $\delta_\text{m}$, is defined as $\delta\rho_\text{m}/\rho_\text{m}$. We remark that, despite the hat notation $(\,\hat{}\,)$ not being adopted, $\Phi$, $\Psi$, $\delta_\text{m}$ and $v$ actually correspond to $\hat{\Phi}(\tau,k_\dagger)$, $\hat{\Psi}(\tau,k_\dagger)$, $\hat{\delta}_{\text{m}}(\tau,k_\dagger)$ and $\hat{v}(\tau,k_\dagger)$, respectively; $Q(\vec{x},k_\dagger)$ has been factored out of Eqs.~(\ref{perturb_5}) and (\ref{deltaR}).

Contrary to what happens in $\Lambda$CDM, $\Phi$ and $\Psi$ are not equal in $f(R)$ gravity \cite{Hwang}:
\begin{equation}
\Psi-\Phi=\frac{f_{RR}\delta R}{f_R}~.
\label{perturb2}
\end{equation}
The above relation follows from the $i$-$j$ component $(i\neq j)$ of Eq.~(\ref{field_eq_pert_fR}). At this stage, we may simplify Eqs.~(\ref{perturb_5}) and (\ref{deltaR}) using the sub-Hubble and quasi-static approximations,\footnote{Sub-Hubble approximation: we assume that the relevant modes are well within the Hubble radius (the `horizon') during the time of interest i.e.~they have $k_\dagger\gg \mathcal{H}$.\\ Quasi-static approximation: can be stated as the condition that $|Y'(\tau)|\lesssim\mathcal{H}|Y|$, where $Y=\Phi$, $\Psi$, $\mathcal{H}$, $\Phi'(\tau)$, $\Psi'(\tau)$ or $\mathcal{H}'(\tau)$ \cite{Tsujikawa2008, Chiu}. In other words, the temporal evolution of $Y$ may essentially be attributed to the expansion of the Universe \cite{Esposito-Farese}, and is thus negligible in comparison to any spatial changes (in $Y$).} whence they become:
\begin{align}
&2f_R\Psi(k_\dagger^2-3\kappa)-f_{RR}k_\dagger^2\delta R+8\pi G a^2 \rho_\text{m}\delta_\text{m}=0~;\label{perturb1B}\\[0.3em]
&\delta R = \frac{2}{a^2}\left[k_\dagger^2(\Phi-2\Psi)+6\kappa\Psi\right]\label{deltaR_B}~.
\end{align}

Let us now take a look at the perturbed version of energy-momentum conservation. The condition $\nabla_\mu T^\mu_\nu=0$ still holds, but in this case $T^\mu_\nu=\tilde{T}^\mu_\nu+\delta T^\mu_\nu$ (the tilde denotes the unperturbed part), and the covariant derivative $\nabla_\mu$ must be constructed from the perturbed metric tensor of Eq.~(\ref{metric-pert}) \cite{Cruz-Dombriz2008}. The relation $\nabla_\mu T^\mu_\nu=0$ may consequently be expanded as follows:
\begin{equation}
\tilde{\nabla}_\mu \tilde{T}^\mu_\nu + \tilde{\nabla}_\mu\delta T^\mu_\nu + \delta \Gamma^\mu_{\sigma\mu}\tilde{T}^\sigma_\nu-\delta \Gamma^\sigma_{\nu\mu}\tilde{T}^\mu_\sigma=0~,
\label{perturbed-cons-eq}
\end{equation}
where $\tilde{\nabla}_\mu$ derives from the background metric, $\tilde{T}^\mu_\nu$ is the unperturbed energy momentum tensor from Eq.~(\ref{energy_momentum}), $\delta T^\mu_\nu$ is its perturbed counterpart [Eq.~(\ref{energy-momentum})], and $\delta\Gamma^\mu_{\nu\sigma}$ represents the perturbed Christoffel symbols. The first term in Eq.~(\ref{perturbed-cons-eq}) has no perturbed components and is therefore conserved separately, giving rise to the familiar relation\footnote{$w_\text{m}$ is the equation-of-state parameter for the matter component [not to be confused with $\omega_\text{m}$ from Eq.~(\ref{change_of_var})].} $\rho_\text{m}'(\tau)=-3\mathcal{H}(\rho_\text{m}+p_\text{m})=-3\mathcal{H}\rho_\text{m}(1+w_\text{m})$. The remaining terms yield the equations \cite{Hwang}:
\begin{align}
\delta_\text{m}'(\tau)&=(1+w_\text{m})\left[-k_\dagger v+3\Psi'(\tau)\right]~;\label{conserv1}\\[0.3em]
v'(\tau)&=\mathcal{H}v(3w_\text{m}-1)+k_\dagger\left(\Phi+\frac{w_\text{m}\delta_\text{m}}{1+w_\text{m}}\right)~,\label{conserv2}
\end{align}
which can readily be combined to give:
\begin{equation}
\delta_\text{m}''(\tau)+\mathcal{H}\delta_\text{m}'(\tau)+k_\dagger^2\Phi-3\Psi''(\tau)-3\mathcal{H}\Psi'(\tau)=0~,
\label{contrast_eq_0}
\end{equation} 
provided that the matter component may be described as a distribution of dust (with $w_\text{m}=0$). The parameter $v$ that appears in Eqs.~(\ref{conserv1}) and (\ref{conserv2}) is the velocity potential associated with $v^i$. In momentum space, we have that\footnote{Every vector can be decomposed into the sum of a scalar part $v_\text{S}^i$ (so called because it may be expressed as the gradient of a scalar field) and a vector part with zero divergence, $v_\text{V}^i$ \cite{Mukhanov, Asgari, Suonio}. In first-order perturbation theory, the two parts evolve independently of each other, and only $v^i_\text{S}$ contributes to the formation of structure \cite{Mukhanov}.} $v^i_\text{S}=-k_\dagger^{-1}\nabla_*^iv$, $\vec{\nabla}_*$ being the covariant derivative operator constructed from the (unperturbed) spatial metric $\gamma_{ij}$ of Eq.~(\ref{metric-pert}). Under the sub-Hubble and quasi-static approximations, Eq.~(\ref{contrast_eq_0}) further simplifies to
\begin{equation}
\delta_\text{m}''(\tau)+\mathcal{H}\delta_\text{m}'(\tau)+k_\dagger^2\Phi=0~.
\label{contrast_eq_1}
\end{equation}
To obtain an expression for $\Phi$, we insert Eq.~(\ref{deltaR_B}) into Eqs.~(\ref{perturb2}) and (\ref{perturb1B}) and solve the last two for $\Phi$ (and $\Psi$), then use the solution to substitute for $\Phi$ in Eq.~(\ref{contrast_eq_1}), which becomes \cite{Tsujikawa}
\begin{equation}
\delta_\text{m}''(\tau)+\mathcal{H}\delta_\text{m}'(\tau)-4\pi \rho_\text{m}\delta_\text{m}a^2G_\text{eff}=0~,\label{main_eq}
\end{equation}
with
\begin{equation}
\frac{G_\text{eff}}{G}=\frac{k_\dagger^2\left[a^2f_R+4f_{RR}(k_\dagger^2-3\kappa)\right]}{f_R\left[3f_{RR}k_\dagger^2\left(k_\dagger^2-4\kappa\right)+a^2f_R\left(k_\dagger^2-3\kappa\right)\right]}~.
\label{Geff_on_G}
\end{equation}

In terms of our new variables [see Eq.~(\ref{change_of_var})], $G_\text{eff}/G$ reads:
\begin{equation}
\frac{G_\text{eff}}{G}=\frac{k_\dagger^2\omega_\text{m} a^3 \eta^2\left[a^2+4\Gamma(k_\dagger^2-3\kappa)\right]}{\Omega_\text{m,0}^\Lambda\left[a^2(k_\dagger^2-3\kappa)+3k_\dagger^2\Gamma(k_\dagger^2-4\kappa)\right]}~.
\end{equation}

\section{Observational data and corresponding likelihoods}
\label{sec:observational}

In this section, we employ Bayesian statistics and place constraints on cosmological/model-specific parameters by utilizing Markov Chain Monte Carlo (MCMC) sampling techniques. We make use of a customized version of the Cosmic Linear Anisotropy Solving System (\textsc{CLASS}) v.2.6.3 \cite{Blas2011}, in conjunction with \textsc{Monte Python} v.3.0.1 \cite{Audren2013, Brinckmann}. For the MCMC part of the study, we consider as baseline parameters the quantities $H_0$ (in units of $\text{km\,s}^{-1}\text{Mpc}^{-1}$), $\Omega_{\mathrm{b},0}\,h^2$, $\Omega_{\mathrm{cdm},0}\,h^2$, $\Omega_{k,0}$, $b$, $n_s$ and $\ln{(10^{10}A_s)}$, where $\Omega_{\mathrm{b},0}$ and $\Omega_{\mathrm{cdm},0}$ are the present-day values of the baryon and cold dark matter density parameters, respectively, $n_s$ stands for the index of the primordial scalar power spectrum and $A_s$ its amplitude, and $h$ is equivalent to $H_0/(100\,\text{km\,s}^{-1}\text{Mpc}^{-1})$. The associated priors are listed in Table~\ref{priors_for_fR}. All other parameters take their \textsc{CLASS} default values, except for the reionization optical depth (which is set to \num{0.0544} \cite{Planck2018VI}). $A_s$ is varied subject to a Gaussian likelihood having a mean of $2.10\times 10^{-9}$ and a standard deviation of $0.03\times 10^{-9}$ \cite{Planck2018VI}. 

The contour plots presented in this work were constructed using the MCMC analysis package \textsc{GetDist} v.1.0.3 \cite{Lewis}.

\renewcommand{\arraystretch}{1.5}
\begin{table}[ht!]
\center
\caption{\label{priors_for_fR} The flat priors assigned to the baseline parameters.}
\vspace{1.5mm}
\begin{tabular}{l c c} 
\hline
\hline
Parameter & Min & Max\\ \hline 
$H_0~\left(\mathrm{km~s}^{-1}\mathrm{Mpc}^{-1}\right)$ & 50 & 90\\ 
$\Omega_{\mathrm{b},0}\,h^2$ & 0.005 & 0.040\\ 
$\Omega_{\mathrm{cdm},0}\,h^2$ & 0.05 & 0.20\\ 
$\Omega_{k,0}$ & -0.3 & 0.3\\
$b$ & 0.0 & 1.0\\  
$n_s$ & 0.75 & 1.25\\
$\ln{(10^{10}A_s)}$ & 2.8 & 3.2\\
\hline
\hline
\end{tabular}
\end{table}

We compare model predictions with measurements of observables related to Type Ia supernovae (SNeIa), the cosmic microwave background (CMB), baryon acoustic oscillations (BAOs), cosmic chronometers and redshift-space distortions (RSDs). Below is a brief description of the respective data sets.

\emph{SNeIa}: We make use of the Pantheon data set, which is based on a sample of 1048 SNeIa in the redshift range $0.01<z<2.3$ \cite{Scolnic}.

\emph{CMB}: Here, we work with four distance priors: the shift parameter $\mathcal{R}$, the acoustic scale $l_\text{A}$, the index of the primordial scalar power spectrum $n_s$, and the quantity $\Omega_{\text{b},0}\,h^2$. The observational values of these four priors and the associated covariance matrix were obtained from Ref.~\cite{Huang2015}. 

\emph{BAO}: Our data set consists of the BAO measurements from the 6dF Galaxy Survey \cite{Beutler2011}, as well as those derived from the main Galaxy sample of SDSS DR7 \cite{Ross2015}, the SDSS-DR12 Ly$\alpha$-quasar cross-correlation function \cite{Bourboux2017} and the Ly$\alpha$-Forest catalogue from the same data release \cite{Bautista2017}, and a BOSS galaxy selection constructed from the CMASS, LOWZ, LOWZE2 and LOWZE3 samples \cite{Alam2017}. This choice of BAO data was made with the aim of removing or at least reducing potential correlations between BAO and RSD measurements. 

\emph{Cosmic chronometers}: The best cosmic chronometers are massive galaxies which acquired most of their stellar mass very rapidly at high redshifts, and have been evolving without major episodes of star formation since then. Consequently, their age may be inferred from that of their stellar population. Once the age difference $\Delta t$ between two such galaxies (located at redshifts $z$ and $z+\Delta z$) has been determined, it is possible to calculate $H(z)$ directly by means of the relation $H(z)=-(1+z)^{-1}\Delta z/\Delta t$ \cite{Jimenez2002, Moresco2012}. We consider the set of Hubble parameter values listed in Table \ref{Tcc}.

$\mathit{H}_0^\mathit{R}$: For part of the analysis, we make use of a Gaussian likelihood for $H_0$ constructed from the local measurement of Ref.~\cite{Riess2019}, which was obtained via the distance-ladder approach and amounts to $74.03\pm1.42\,\text{km\,s}^{-1}\text{Mpc}^{-1}$. The tension between this value of the Hubble constant and the \emph{Planck} constraints on $H_0$ \cite{Planck2018VI} is still unresolved, so it is crucial that we investigate any implications that an $H_0$ likelihood might have for the inferred $f(R)$ parameter constraints. 

{
\renewcommand{\arraystretch}{1.2}
\begin{table}[h]
\setcounter{mpfootnote}{\value{footnote}}
\renewcommand{\thempfootnote}{\arabic{mpfootnote}}
\caption{\label{Tcc} Cosmic chronometer data. Each value of $H(z)$ is listed together with the corresponding redshift $z$ and error $\sigma$.}
\vspace{1.5mm}
\begin{tabular}{@{\hskip 0.2cm}c@{\hskip 0.5cm} S[table-format=2.5]@{\hskip 0.6cm} S[table-format=3.2]@{\hskip 0.5cm} S[table-format=3.2]}
\hline
\hline
Ref.  & {$z$} & {$~H(z)$} & {$\sigma$}\\
~ & ~ & \multicolumn{2}{l}{$\left(\si{km.s^{-1}.Mpc^{-1}}\right)$}\\
\hline
\cite{Zhang2014} & 0.0700 & 69.0 & 19.6\\
\cite{Zhang2014} & 0.1200 & 68.6 & 26.2\\
\cite{Simon2005} & 0.1700 & 83.0 & 8.0\\
\cite{Moresco2012} & 0.1791 & 75.0 &  4.0\\
\cite{Moresco2012} & 0.1993 & 75.0 & 5.0\\
\cite{Zhang2014} & 0.2000 & 72.9 & 29.6\\
\cite{Simon2005} & 0.2700 & 77.0 & 14.0\\
\cite{Zhang2014} & 0.2800 & 88.8 & 36.6\\ 
\cite{Moresco2012} & 0.3519 & 83.0 & 14.0\\
\cite{Moresco2016} & 0.3802 & 83.0 & 13.6\\
\cite{Simon2005} & 0.4000 & 95.0 & 17.0\\
\cite{Moresco2016} & 0.4004 & 77.0 & 10.2\\
\cite{Moresco2016} & 0.4247 & 87.1 & 11.2\\
\cite{Moresco2016} & 0.4497 & 92.8 & 12.9\\
\cite{Ratsimbazafy2017}\footnotemark[7] & 0.4700 & 89.0 & 49.6\\
\cite{Moresco2016} & 0.4783 & 80.9 & 9.0\\
\cite{Stern2010} & 0.4800 & 97.0 & 62.0\\
\cite{Moresco2012} & 0.5929 & 104.0 & 13.0\\
\cite{Moresco2012} & 0.6797 & 92.0 & 8.0\\
\cite{Moresco2012} & 0.7812 & 105.0 & 12.0\\
\cite{Moresco2012} & 0.8754 & 125.0 & 17.0\\
\cite{Stern2010} & 0.8800 & 90.0 & 40.0\\
\cite{Simon2005} & 0.9000 & 117.0 & 23.0\\
\cite{Moresco2012} & 1.0370 & 154.0 & 20.0\\
\cite{Simon2005} & 1.3000 & 168.0 & 17.0\\
\cite{Moresco2015} & 1.3630 & 160.0 & 33.6\\
\cite{Simon2005} & 1.4300 & 177.0 & 18.0\\
\cite{Simon2005} & 1.5300 & 140.0 & 14.0\\
\cite{Simon2005} & 1.7500 & 202.0 & 40.0\\
\cite{Moresco2015} & 1.9650 & 186.5 & 50.4\\
\hline
\hline
\end{tabular}
\end{table}
}

\footnotetext[7]{In this case, $\sigma$ was calculated by summing the systematic and statistical errors in quadrature.}
\addtocounter{footnote}{1}
{
\renewcommand{\arraystretch}{1.3}
\begin{table}[ht!]
\centering
\caption{\label{TLSS}LSS data. Each $f\sigma_8(z)$ measurement is listed together with the corresponding redshift $z$ and error $\sigma_{f\sigma_8}^{}$, while Column 5 contains the values of $\Omega_\text{m,0}$ for the respective fiducial cosmologies.} 
\vspace{1.5mm}
\begin{tabular}{c S[table-format=2.3] S[table-format=2.4] S[table-format=2.4] S[table-format=2.3]} 
\hline
\hline
Ref. & {$z$} & {$f\sigma_8(z)$} & {$\sigma_{f\sigma_8}^{}$} & {$\Omega_{\text{m},0}^\text{fid}$}\\ 
\hline
\cite{Davis, Hudson} & 0.020 & 0.314 & 0.048 & 0.266\\
\cite{Blake2012} & 0.440 & 0.413 & 0.080 & 0.270\\
\cite{Blake2012} & 0.600 & 0.390 & 0.063 & 0.270\\
\cite{Blake2012} & 0.730 & 0.437 & 0.072 & 0.270\\
\cite{Pezzotta} & 0.600 & 0.550 & 0.120 & 0.300\\
\cite{Pezzotta} & 0.860 & 0.400 & 0.110 & 0.300\\
\cite{Okumura} & 1.400 & 0.482 & 0.116 & 0.270\\
\cite{Zhao} & 0.978 & 0.379 & 0.176 & 0.310\\
\cite{Zhao} & 1.230 & 0.385 & 0.099 & 0.310\\
\cite{Zhao} & 1.526 & 0.342 & 0.070 & 0.310\\
\cite{Zhao} & 1.944 & 0.364 & 0.106 & 0.310\\
\hline
\hline
\end{tabular}
\end{table}
}

\emph{RSD}: The collection of $f\sigma_8(z)$ data points we work with is provided in Table \ref{TLSS}, and has been adopted from the compilation in Ref.~\cite{Sagredo}. 

To calculate $f\sigma_8(z)$ for a given model, we need two quantities: the growth rate $f(z)$, and the standard deviation of density perturbations in spheres of radius $8\,h^{-1}\,\text{Mpc}$, $\sigma_8(z)$. The former is a function of $\delta_\text{m}$, the matter density contrast:
\begin{equation}
f=\frac{\text{d}(\ln{\delta_\text{m}})}{\text{d}\ln{a}}~.
\end{equation}
$\delta_\text{m}$ may in turn be obtained from Eq.~(\ref{main_eq}) by solving it as part of the system of differential equations given by Eqs. (\ref{etaprime})--(\ref{ommprime}) [with $\omega_\text{r}$ and $K$ as defined in Eq.~(\ref{change_of_var})].\footnote{We solve everything in terms of $a$ by making use of the relations $\text{d}/\text{d}z = -a^2\,\text{d}/\text{d}a$ and $\text{d}/\text{d}\tau = Ha^2\,\text{d}/\text{d}a$.} 

Since the expression for $G_\text{eff}/G$ [Eq.~(\ref{Geff_on_G})] is dependent on $k_\dagger$, it is necessary to choose an appropriate comoving wave number at which to evaluate $f(z)$. We shall focus exclusively on values of $k_\dagger$ in the range $0.02\,h \,\text{Mpc}^{-1}\leq k_\dagger\leq 0.2\,h\, \text{Mpc}^{-1}$. This choice of bounds is based on two considerations: firstly, modes with $k_\dagger\lesssim 0.2\,h\, \text{Mpc}^{-1}$ represent perturbations which may safely be considered linear \cite{Dodelson2003}, and secondly, the mode that crosses the horizon at matter-radiation equality has\footnote{As calculated for a $\Lambda$CDM cosmology by using \emph{Planck} values \cite{Planck2018VI}.} $k_\dagger\sim 0.015\,h\,\text{Mpc}^{-1}$. The latter implies that smaller scales -- corresponding to a larger $k_\dagger$ -- would be well within the horizon during the epochs of interest i.e.~deep in the matter era and throughout the subsequent period of acceleration. We shall therefore determine $f(z)$ for two different values of $k_\dagger$: $k_\dagger=0.1\,h\,\text{Mpc}^{-1}$ and $k_\dagger=0.05\,h\,\text{Mpc}^{-1}$. In both cases, $k_\dagger^2$ is much larger than we could reasonably expect $|\kappa|$ to be $(k_\dagger^2\gg |\kappa|)$, and so Eq.~(\ref{Geff_on_G}) may be simplified to:
\begin{equation}
\frac{G_\text{eff}}{G}=\frac{a^2f_R+4f_{RR}k_\dagger^2}{f_R\left(3f_{RR}k_\dagger^2+a^2f_R\right)}~.
\end{equation}

Finally, to solve Eq.~(\ref{main_eq}) we require a pair of initial conditions. The simplest choice is $\delta_\text{m}(a_\text{ini})=a_\text{ini}$ and $\delta_\text{m}'(a=a_\text{ini})=1$ [or equivalently, $\delta_\text{m}(\tau_\text{ini})=a_\text{ini}$ and $\delta_\text{m}'(\tau=\tau_\text{ini}^{})=H(a_\text{ini}^{}) a_\text{ini}^2$\,], $a_\text{ini}^{}$ and $\tau_\text{ini}^{}$ being the scale factor and conformal time, respectively, at which the initial conditions are applied. However, this is only an option if two requirements are met: firstly, $a_\text{ini}$ must correspond to a time when the Universe is deep in the matter-dominated epoch. Secondly, the given initial conditions are only valid for a $\Lambda$CDM cosmology \cite{Dodelson2003}. This is where the quantity $z_\text{bound}$ derived in subsection~\ref{subsec:fR_models} turns out to be useful. Let us recall that when $z=z_\text{bound}$, $f(R)=R-2\Lambda(1-\epsilon)$ for some $\epsilon \ll 1$. Therefore, by calculating $z_\text{bound}$ for a given set of values $\{b,\,\Omega_{\text{m},0},\,\Omega_{\Lambda,0}\}$, we can ensure that the initial conditions in question are applied during the `$\Lambda$CDM epoch', i.e.~at a redshift $z_\text{ini}\,(=a_\text{ini}^{-1}-1)$ which satisfies $z_\text{ini}\geq z_\text{bound}$. This is also important because $z_\text{ini}$ is the redshift at which we stop providing the integrator with the equations for $\Lambda$CDM and switch to $f(R)$ [we check that the value of $\Gamma$ exceeds machine precision before applying Eq.~(\ref{xprime})].

Since we want $z_\text{ini}$ to correspond to the matter-dominated epoch, the specific value of $\epsilon$ is dependent upon the dynamics of the model being considered. For instance, in the case of the Hu-Sawicki model, it suffices to have $\epsilon\sim 10^{-5}$. On the other hand, the Exponential model converges to $\Lambda$CDM extremely rapidly and hence a smaller $\epsilon$ $(\sim 10^{-50})$ works better.\footnote{In the case of the Exponential and Tsujikawa models, the convergence is so fast that $\delta_\text{ini}^{}$ and $\delta'_\text{ini}$ must be applied at a redshift $z_\text{ini}$ which is strictly greater than $z_\text{bound}$ (if the condition of matter domination is to be met). To avoid computations with very small numbers, we switch from $\Lambda$CDM to $f(R)$ at a redshift $z$ in the range $z_\text{ini}<z<z_\text{bound}$.} It should be noted, however, that the final results are \emph{not} dependent on the exact value of $\epsilon$ (provided $\epsilon$ is sufficiently small). This was verified for both the Exponential and Tsujikawa models.

\begin{figure}[t!]
\begin{center}
\includegraphics[width=0.95\columnwidth]{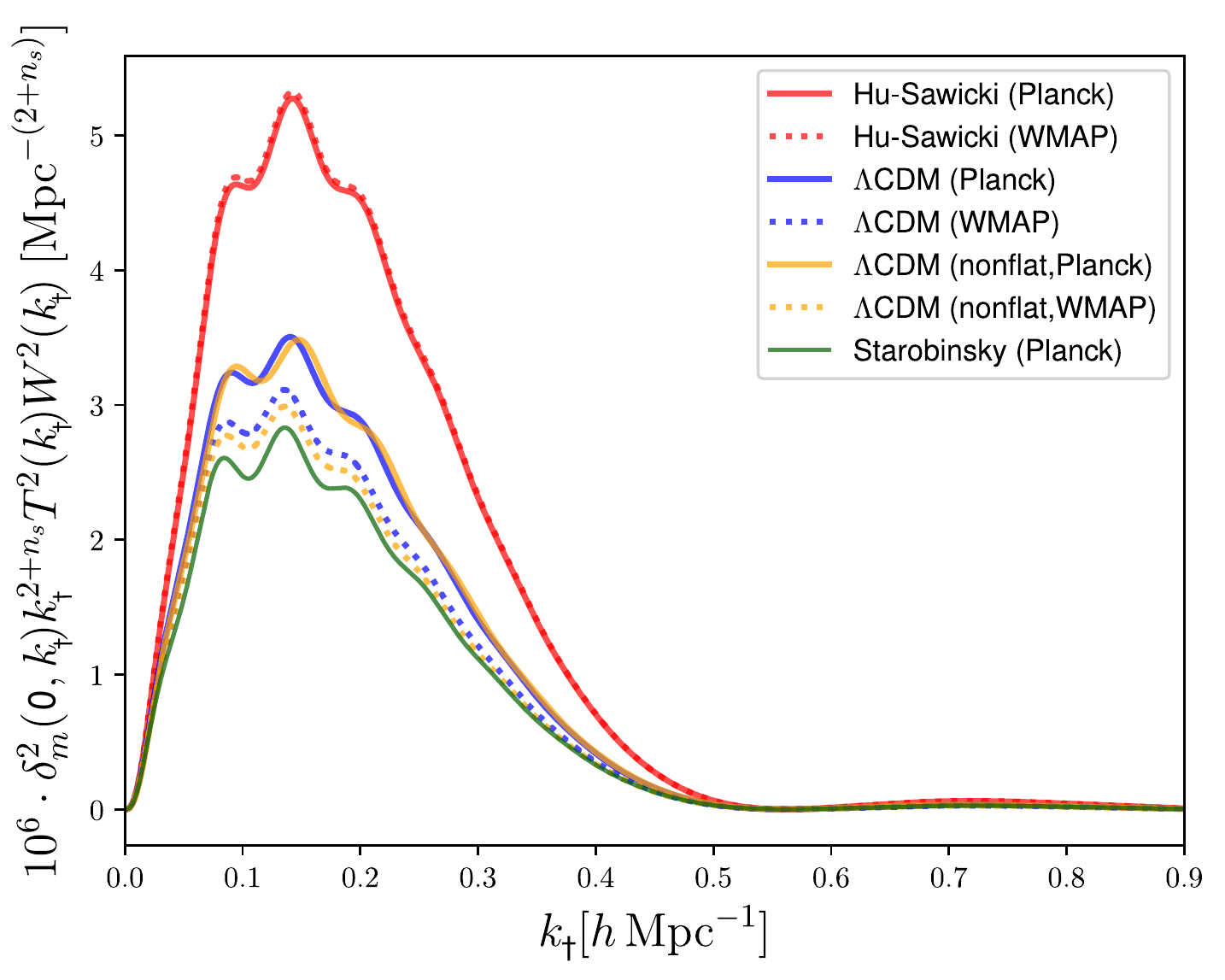}
\caption{\label{integrand_for_fsig8}{The variation of the integrand in Eq.~(\ref{sigma_8_z}) with comoving wave number at $z=0$ (only the $k_\dagger$-dependent part is plotted). The curves labeled `Hu-Sawicki' (`Starobinsky') are based on the mean parameter values presented in Ref.~\cite{Romero} (Ref.~\cite{Sultana2019}), with the remaining parameters fixed according to either the \emph{Planck} 2018 \cite{Planck2018VI} or the WMAP 9-year \cite{WMAP} results. The label `$\Lambda$CDM' indicates that only \emph{Planck} or WMAP values were used.}}
\end{center}
\end{figure}

As previously suggested, the model is allowed to evolve identically to $\Lambda$CDM for redshifts $z>z_\text{ini}$. At $z=z_\text{ini}$, therefore, we must also specify initial conditions for Eqs.~(\ref{etaprime})--(\ref{ommprime}). These conditions are essentially the $\Lambda$CDM limits of $\eta$ and the parameters defined in Eq.~(\ref{change_of_var}).

\renewcommand{\arraystretch}{1.5}
\begin{table*}[ht]
\center
\caption{\label{T_HS} Mean values and $1\sigma$ confidence intervals for the Hu-Sawicki model parameters.}
\vspace{1em}
\begin{tabular}{l@{\hskip 1.0cm} A@{\hskip 1.0cm} A@{\hskip 1.0cm} A@{\hskip 1.0cm} A}  
\hline
\hline
Parameter & \multicolumn{2}{c}{Flat~~~~} & \multicolumn{2}{c}{Nonflat~~~~} & \multicolumn{2}{c}{Nonflat ($+H_0^\text{R}$)} & \multicolumn{2}{c}{Nonflat}\\ 
~~ & \multicolumn{2}{c}{$k_\dagger=0.1\,h\,\text{Mpc}^{-1}$} & \multicolumn{2}{c}{$k_\dagger=0.1\,h\,\text{Mpc}^{-1}$} & \multicolumn{2}{c}{$k_\dagger=0.1\,h\,\text{Mpc}^{-1}$} & \multicolumn{2}{c}{$k_\dagger=0.05\,h\,\text{Mpc}^{-1}$} \\ \hline
$H_0$&68.5280&_{-0.4851}^{+0.4797}&68.6490&_{-0.6645}^{+0.6385}&69.6890&_{-0.6139}^{+0.6044}&68.6880&_{-0.6640}^{+0.6543}\\ 
$10^3\Omega_{\mathrm{b},0}^{}h^2$&22.4790&_{-0.1460}^{+0.1441}&22.4290&_{-0.1714}^{+0.1727}&22.4500&_{-0.1722}^{+0.1723}&22.4310&_{-0.1808}^{+0.1776}\\ 
$\Omega_{\mathrm{cdm},0}^{}h^2$&0.1188&_{-0.0010}^{+0.0010}&0.1195&_{-0.0016}^{+0.0016}&0.1195&_{-0.0016}^{+0.0015}&0.1194&_{-0.0017}^{+0.0016}\\ 
$10^3\Omega_{k,0}$&-&-&0.8322&_{-1.9341}^{+1.9912}&2.7676&_{-1.8470}^{+1.8587}&0.8799&_{-2.0515}^{+2.0596}\\ 
$10^4 b$&0.2739&_{-0.2739}^{+0.1789}&0.7607&_{-0.7607}^{+0.5798}&0.2089&_{-0.2089}^{+0.0343}&1.9176&_{-1.9176}^{+8.3473}\\ 
$n_s$&0.9705&_{-0.0041}^{+0.0041}&0.9688&_{-0.0051}^{+0.0051}&0.9689&_{-0.0051}^{+0.0052}&0.9689&_{-0.0053}^{+0.0054}\\ 
$\ln{\left(10^{10}A_s\right)}$&3.0442&_{-0.0144}^{+0.0148}&3.0443&_{-0.0145}^{+0.0148}&3.0442&_{-0.0146}^{+0.0148}&3.0441&_{-0.0145}^{+0.0148}\\ \hline
$\Omega_{\Lambda,0}$\,\emph{(derived)}&0.6989&_{-0.0061}^{+0.0062}&0.6979&_{-0.0062}^{+0.0062}&0.7049&_{-0.0057}^{+0.0059}&0.6982&_{-0.0065}^{+0.0067}\\
\hline
\hline
\end{tabular}
\end{table*}

The second quantity we need in order to determine $f\sigma_8(z)$ is the standard deviation, $\sigma_8(z)$, which is given by:
\begin{align}
&\sigma_8^2(z) =\notag\\&\int\limits_0^\infty \delta_\text{m}^2(z,k_\dagger)\,k_\dagger^{2+n_s}\left[\frac{4A_s\,k_*^{1-n_s}}{25H_0^4\Omega_{\text{m},0}^2}\right]T^2(k_\dagger)W^2(k_\dagger)\,\text{d}k_\dagger~.
\label{sigma_8_z}
\end{align}
In the above, $k_*$ denotes the pivot scale at which $n_s$ and $A_s$ are defined (here equal to $0.05\,\text{Mpc}^{-1}$ \cite{Planck2018VI}), and the function $W(k_\dagger)$ represents the Fourier transform of a spherical top-hat window function having radius $R_8$ $(=8\,h^{-1}\text{Mpc})$:
\begin{equation}
W(k_\dagger) = \frac{3}{k_\dagger^2R_8^2}\left[\frac{\sin{(k_\dagger R_8)}}{k_\dagger R_8}-\cos{(k_\dagger R_8)}\right]~.
\end{equation}

We model the transfer function $T(k_\dagger)$ as detailed in the work of Eisenstein and Hu \cite{Eisenstein1998}.

Since the density contrast function is not scale invariant in $f(R)$ gravity, every different value of $k_\dagger$ we consider entails that we solve Eq.~(\ref{main_eq}) numerically for $\delta_\text{m}$ over the required redshift range. The results are stored in a table, and at a given redshift $z$, $\delta_\text{m}(z,k_\dagger)$ is extracted from the table by interpolation for all relevant comoving wave numbers $k_\dagger$, and used to calculate the integrand in Eq.~(\ref{sigma_8_z}). 
Hence it becomes necessary to truncate the range of wave numbers over which integration is performed. To this end, we plot the integrand as a function of $k_\dagger$ for various values of $z$, then infer the cut-off point from the outcome. Fig.~\ref{integrand_for_fsig8} shows a sample of such plots. The intuitive choice would be $k_\dagger\approx0.5\,h\,\text{Mpc}^{-1}$, but one must keep in mind that perturbations evolve non-linearly for values of $k_\dagger$ larger than around $0.2\,h\,\text{Mpc}^{-1}$. Consequently, the linear perturbation equations we work with do not give an accurate description of structure growth beyond this limit. Nonetheless, $k_\dagger=0.5\,h\,\text{Mpc}^{-1}$ is a good starting point. We refine it further by calculating $\sigma_{8,0}$ [$=\sigma_8(z=0)$] for a $\Lambda$CDM cosmology from Eq.~(\ref{sigma_8_z}) and comparing it with the value returned by the default \textsc{CLASS} code. The two are closest when the upper integration limit is $\approx 0.4\,h\,\text{Mpc}^{-1}$. 

Finally, we correct for the Alcock-Paczynski effect by scaling (multiplying) the theoretical value of $f\sigma_8(z)$ by the factor $H_\text{fid}(z)\,d_\text{A,fid}(z)/H(z) d_\text{A}(z)$ \cite{Macaulay,Lavrentios}, where $d_\text{A}(z)$ is the angular diameter distance to redshift $z$ and a subscript `fid' labels quantities pertaining to the fiducial cosmology. This refers to the cosmology in whose framework the \emph{observational} value at the same redshift, $[f\sigma_8(z)]_\text{obs}$, was obtained. 

\section{Results}
\label{sec:results}
\subsubsection{The Hu-Sawicki model}

\begin{figure}[ht]
    \begin{center}
        \includegraphics[width=\columnwidth]{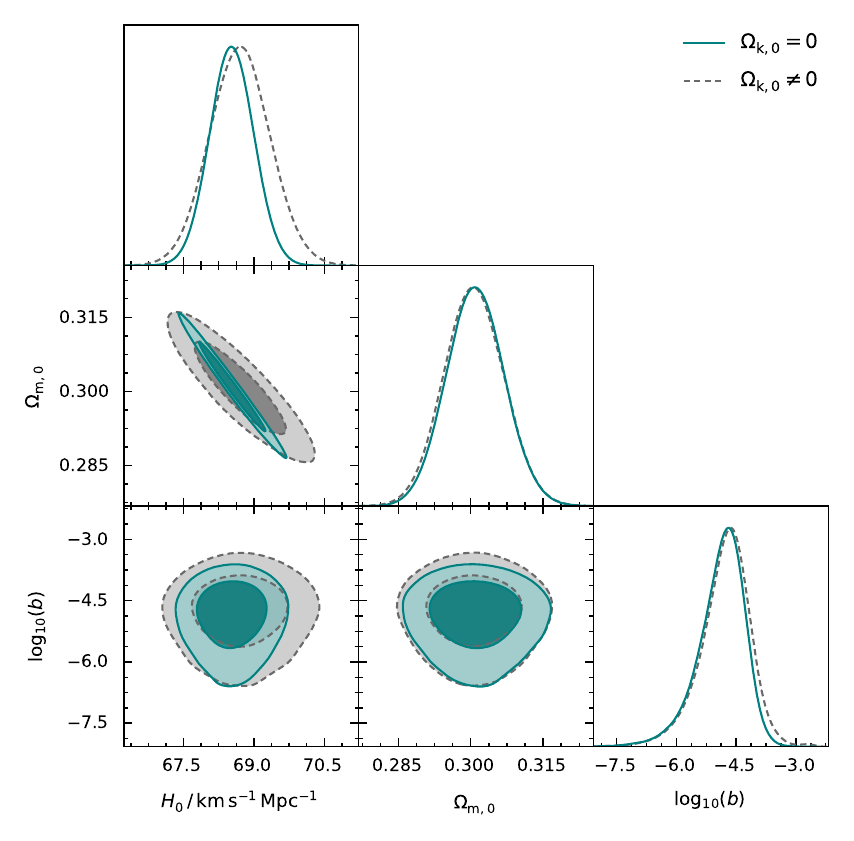}
        \caption{\label{fig:HS_flat_triangle}{A comparison between the posterior distributions for the main parameters of the spatially flat $(\Omega_{k,0}=0)$ Hu-Sawicki model and its non-flat $(\Omega_{k,0}\neq0)$ counterpart. Darker (lighter) shades denote $1\sigma$ ($2\sigma$) confidence regions.}}
    \end{center}
\end{figure}

\begin{figure*}[ht]
    \begin{center}
        \includegraphics[width=0.57\paperwidth]{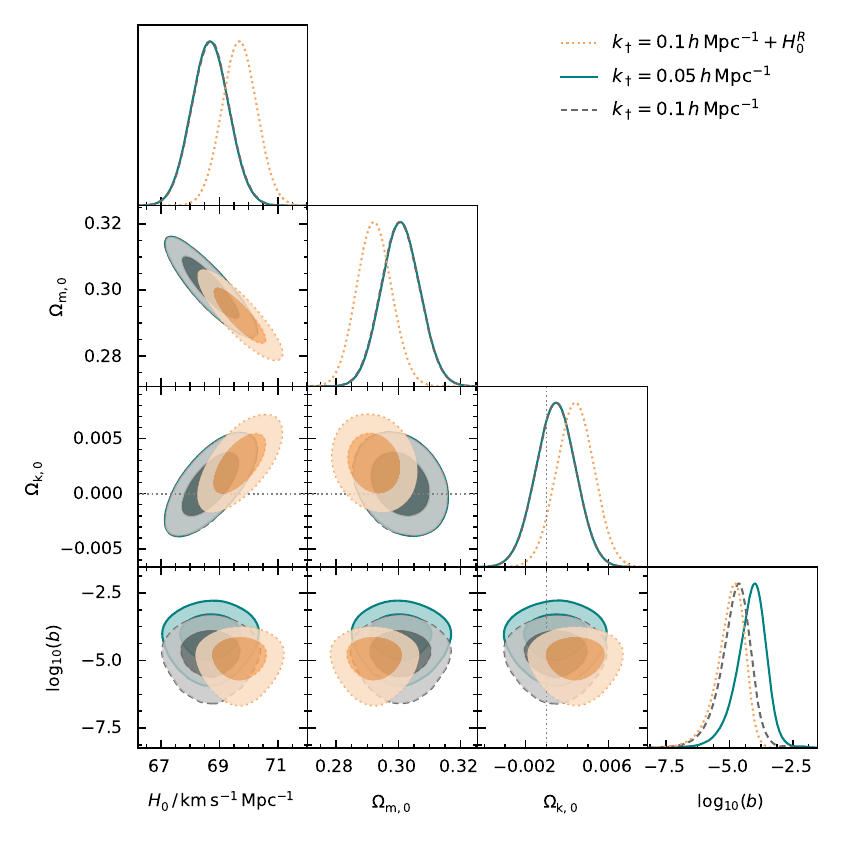}
        \caption{\label{fig:HS_nonflat_triangle}{Marginalized 2D and 1D posterior distributions for the main parameters of the spatially non-flat Hu-Sawicki model. The `$+H_0^\text{R}$' that appears in the legend indicates the use of a Gaussian likelihood for $H_0$.}}
    \end{center}
\end{figure*}

~~~~~\\
The mean values and $1\sigma$ confidence intervals for the Hu-Sawicki model are presented in Table \ref{T_HS}, with a selection of 2D and 1D posterior distributions shown in Figs.~\ref{fig:HS_flat_triangle} and \ref{fig:HS_nonflat_triangle}. The tabulated results and the latter figure allow us to deduce that varying $k_\dagger$ from $0.1\,h\,\text{Mpc}^{-1}$ to $0.05\,h\,\text{Mpc}^{-1}$ has minimal impact on the cosmological parameter constraints, but shifts the mean deviation parameter $b$ to larger values. Another prominent characteristic is the negative correlation between $\Omega_{\text{m},0}$ and $H_0$. This is due to the fact that the theoretical expressions for many of the observables we use contain the product $\Omega_{\text{m},0}^{}\,H_0^2$, which implies that combinations of the parameters $\Omega_{\text{m},0}$ and $H_0$ that give rise to the same value of $\Omega_{\text{m},0}^{}\,H_0^2$ are equally likely with respect to the observables in question. Similarly, the effects that variations in $\Omega_{k,0}$ would have on the primary CMB anisotropies may be offset by changes in $H_0$. This happens if the angular diameter distance to last scattering remains unaltered \cite{Planck2015XIII}, and explains the correlation between $\Omega_{k,0}$ and $H_0$.

The addition of $H_0^\text{R}$ to the data set yields a larger mean value of $H_0$, as expected. In view of the above-mentioned correlations, it becomes clear why this also causes a marked shift towards smaller (larger) values in the 1D posterior distribution for $\Omega_{\text{m},0}$ ($\Omega_{k,0}$).  

Fig.~\ref{fig:HS_flat_triangle} illustrates how constraints are altered when the assumption of spatial flatness is relaxed. On the whole, confidence intervals tend to become wider,\footnote{This is easier to deduce from the tabulated results.} with the difference being most apparent for the $\Omega_{\text{m},0}$ vs $H_0$ contour plot. Indeed, the negative correlation between the two quantities is much stronger for a flat universe. This feature arises due to the fact that $H_0$ is correlated with $\Omega_{k,0}$.


\begin{figure}[ht]
    \begin{center}
        \includegraphics[width=\columnwidth]{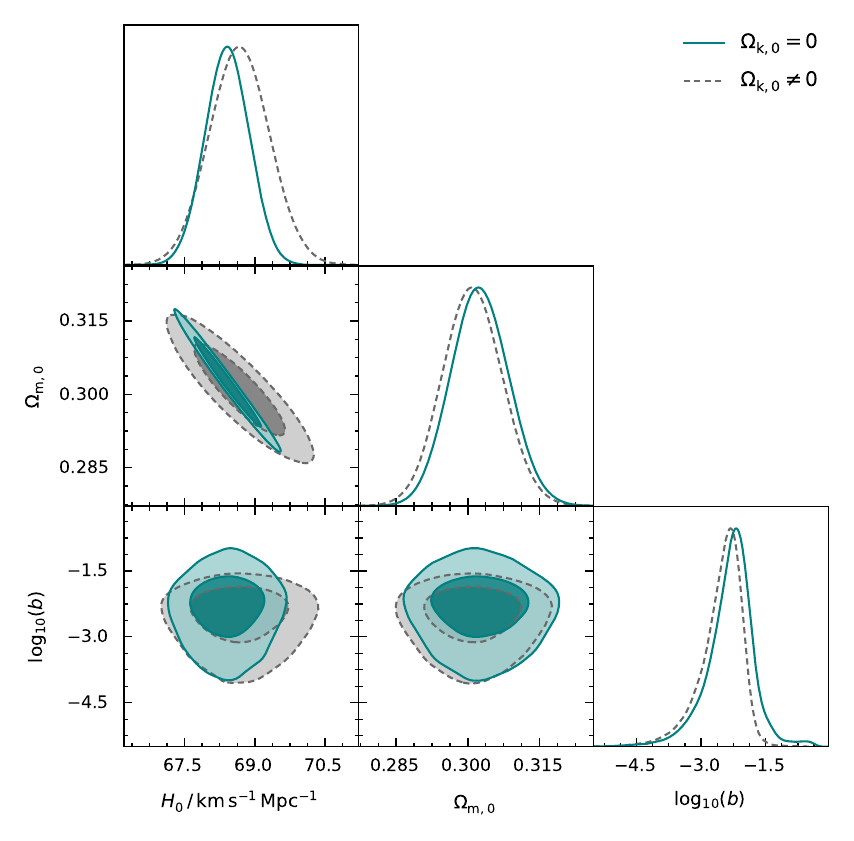}
        \caption{\label{fig:Star_flat_triangle}{A comparison between the posterior distributions for the main parameters of the spatially flat $(\Omega_{k,0}=0)$ Starobinsky model and its non-flat $(\Omega_{k,0}\neq0)$ counterpart.}}
    \end{center}
\end{figure}

\subsubsection{The Starobinsky model}

\renewcommand{\arraystretch}{1.5}
\begin{table*}[ht]
\center
\caption{\label{T_Staro} Mean values and $1\sigma$ confidence intervals for the Starobinsky model parameters.}
\vspace{1.5mm}
\begin{tabular}{l@{\hskip 1.0cm} A@{\hskip 1.0cm} A@{\hskip 1.0cm} A@{\hskip 1.0cm} A}  
\hline
\hline
Parameter & \multicolumn{2}{c}{Flat~~~~} & \multicolumn{2}{c}{Nonflat~~~~} & \multicolumn{2}{c}{Nonflat ($+H_0^\text{R}$)} & \multicolumn{2}{c}{Nonflat}\\ 
~~ & \multicolumn{2}{c}{$k_\dagger=0.1\,h\,\text{Mpc}^{-1}$} & \multicolumn{2}{c}{$k_\dagger=0.1\,h\,\text{Mpc}^{-1}$} & \multicolumn{2}{c}{$k_\dagger=0.1\,h\,\text{Mpc}^{-1}$} & \multicolumn{2}{c}{$k_\dagger=0.05\,h\,\text{Mpc}^{-1}$} \\ \hline
$H_0$&68.4080&_{-0.4852}^{+0.4792}&68.6270&_{-0.6569}^{+0.6541}&69.5520&_{-0.6178}^{+0.6097}&68.5910&_{-0.6586}^{+0.6512}\\ 
$10^3\Omega_{\mathrm{b},0}^{}h^2$&22.4560&_{-0.1442}^{+0.1451}&22.4260&_{-0.1734}^{+0.1700}&22.4390&_{-0.1718}^{+0.1716}&22.4250&_{-0.1729}^{+0.1703}\\ 
$\Omega_{\mathrm{cdm},0}^{}h^2$&0.1191&_{-0.0010}^{+0.0010}&0.1195&_{-0.0015}^{+0.0015}&0.1196&_{-0.0016}^{+0.0015}&0.1195&_{-0.0015}^{+0.0015}\\ 
$10^3\Omega_{k,0}$&-&-&0.8160&_{-1.9055}^{+1.9832}&2.6443&_{-1.8506}^{+1.8679}&0.7448&_{-1.8984}^{+2.0605}\\ 
$b$&0.0122&_{-0.0122}^{+0.0055}&0.0057&_{-0.0057}^{+0.0011}&0.0064&_{-0.0064}^{+0.0074}&0.0132&_{-0.0132}^{+0.0021}\\ 
$n_s$&0.9699&_{-0.0041}^{+0.0041}&0.9687&_{-0.0051}^{+0.0051}&0.9686&_{-0.0051}^{+0.0052}&0.9687&_{-0.0051}^{+0.0050}\\ 
$\ln{\left(10^{10}A_s\right)}$&3.0441&_{-0.0144}^{+0.0149}&3.0443&_{-0.0143}^{+0.0149}&3.0442&_{-0.0145}^{+0.0149}&3.0442&_{-0.0144}^{+0.0152}\\ \hline
$\Omega_{\Lambda,0}\,\emph{(derived)} $&0.6974&_{-0.0060}^{+0.0063}&0.6976&_{-0.0061}^{+0.0063}&0.7036&_{-0.0057}^{+0.0059}&0.6974&_{-0.0058}^{+0.0064}\\
\hline
\hline
\end{tabular}
\end{table*}

\renewcommand{\arraystretch}{1.5}
\begin{table*}[ht]
\center
\caption{\label{T_Exp} Mean values and $1\sigma$ confidence intervals for the Exponential model parameters.}
\vspace{1.5mm}
\begin{tabular}{l@{\hskip 1.0cm} A@{\hskip 1.0cm} A@{\hskip 1.0cm} A@{\hskip 1.0cm} A}  
\hline
\hline
Parameter & \multicolumn{2}{c}{Flat~~~~} & \multicolumn{2}{c}{Nonflat~~~~} & \multicolumn{2}{c}{Nonflat ($+H_0^\text{R}$)} & \multicolumn{2}{c}{Nonflat}\\ 
~~ & \multicolumn{2}{c}{$k_\dagger=0.1\,h\,\text{Mpc}^{-1}$} & \multicolumn{2}{c}{$k_\dagger=0.1\,h\,\text{Mpc}^{-1}$} & \multicolumn{2}{c}{$k_\dagger=0.1\,h\,\text{Mpc}^{-1}$} & \multicolumn{2}{c}{$k_\dagger=0.05\,h\,\text{Mpc}^{-1}$} \\ \hline
$H_0$&68.3990&_{-0.4810}^{+0.4800}&68.5510&_{-0.6648}^{+0.6598}&69.5060&_{-0.6096}^{+0.6030}&68.5650&_{-0.6657}^{+0.6520}\\ 
$10^3\Omega_{\mathrm{b},0}^{} h^2$&22.4560&_{-0.1448}^{+0.1457}&22.4210&_{-0.1719}^{+0.1701}&22.4360&_{-0.1697}^{+0.1702}&22.4220&_{-0.1715}^{+0.1719}\\ 
$\Omega_{\mathrm{cdm},0}^{} h^2$&0.1191&_{-0.0010}^{+0.0010}&0.1195&_{-0.0016}^{+0.0016}&0.1196&_{-0.0016}^{+0.0015}&0.1195&_{-0.0016}^{+0.0015}\\ 
$10^3\Omega_{k,0}$&-&-&0.7000&_{-1.9734}^{+2.0506}&2.5938&_{-1.8318}^{+1.8411}&0.7216&_{-1.9544}^{+1.9878}\\ 
$b$ & 0.1601&_{-0.1565}^{+0.0916} & 0.1641&_{-0.1599}^{+0.0850} & 0.1608&_{-0.1800}^{+0.1800}& 0.1769&_{-0.1727}^{+0.0920}\\ 
$n_s$&0.9697&_{-0.0041}^{+0.0041}&0.9686&_{-0.0052}^{+0.0051}&0.9684&_{-0.0051}^{+0.0051}&0.9686&_{-0.0052}^{+0.0051}\\ 
$\ln{\left(10^{10}A_s\right)}$&3.0443&_{-0.0146}^{+0.0147}&3.0441&_{-0.0144}^{+0.0149}&3.0440&_{-0.0145}^{+0.0148}&3.0442&_{-0.0144}^{+0.0149}\\ \hline
$\Omega_{\Lambda,0}$\,\emph{(derived)}&0.6972&_{-0.0061}^{+0.0063}&0.6971&_{-0.0061}^{+0.0063}&0.7032&_{-0.0057}^{+0.0060}&0.6971&_{-0.0061}^{+0.0063}\\
\hline
\hline
\end{tabular}
\end{table*}

\begin{figure*}[ht]
    \begin{center}
        \includegraphics[width=0.55\paperwidth]{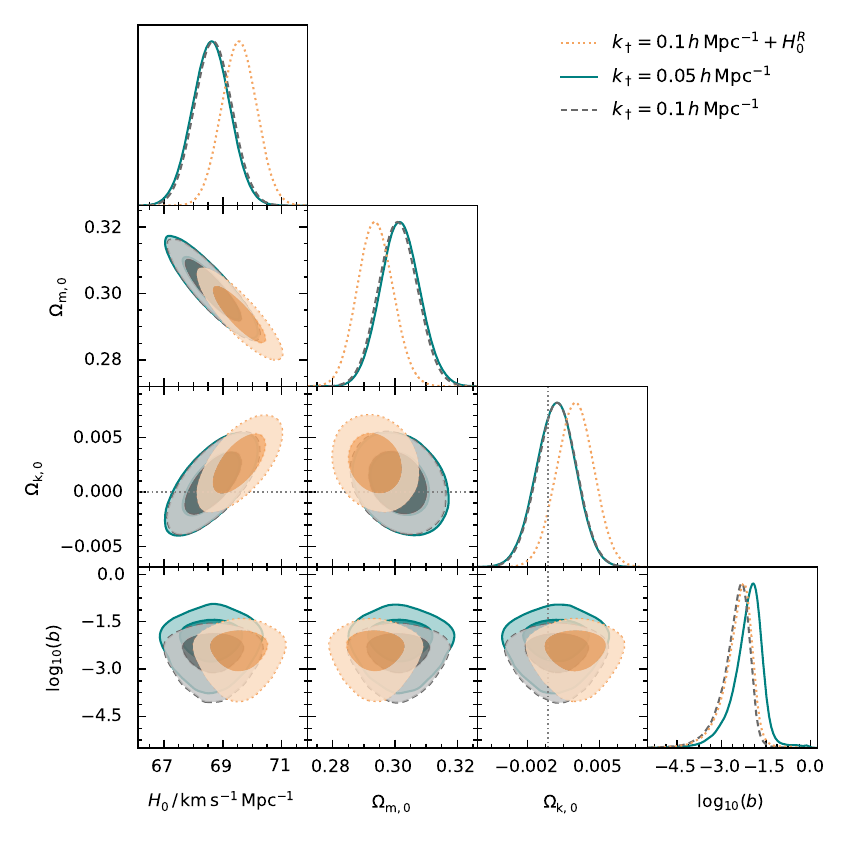}
        \caption{\label{fig:Star_nonflat_triangle}{Marginalized 2D and 1D posterior distributions for the main parameters of the spatially non-flat Starobinsky model.}}
    \end{center}
\end{figure*}

The inferred parameter constraints, including those for the deviation parameter, are presented in Table \ref{T_Staro} and in Figs.~\ref{fig:Star_flat_triangle} and \ref{fig:Star_nonflat_triangle}. The effects that using a different wave number, introducing curvature or including a Gaussian likelihood for $H_0$ have on the constraints are much the same as for the Hu-Sawicki model. We again note that the mean value of $b$ becomes slightly larger when $k_\dagger=0.05\,h\,\mathrm{Mpc}^{-1}$ (relative to what we get when $k_\dagger=0.1\,h\,\mathrm{Mpc}^{-1}$ in the non-flat case), and that the $H_0^\text{R}$ likelihood makes the constraints on $\Omega_{k,0}$ compatible with a spatially curved universe at $\sim1\sigma$.


\subsubsection{The Exponential model}

\begin{figure}[ht]
    \begin{center}
        \includegraphics[width=\columnwidth]{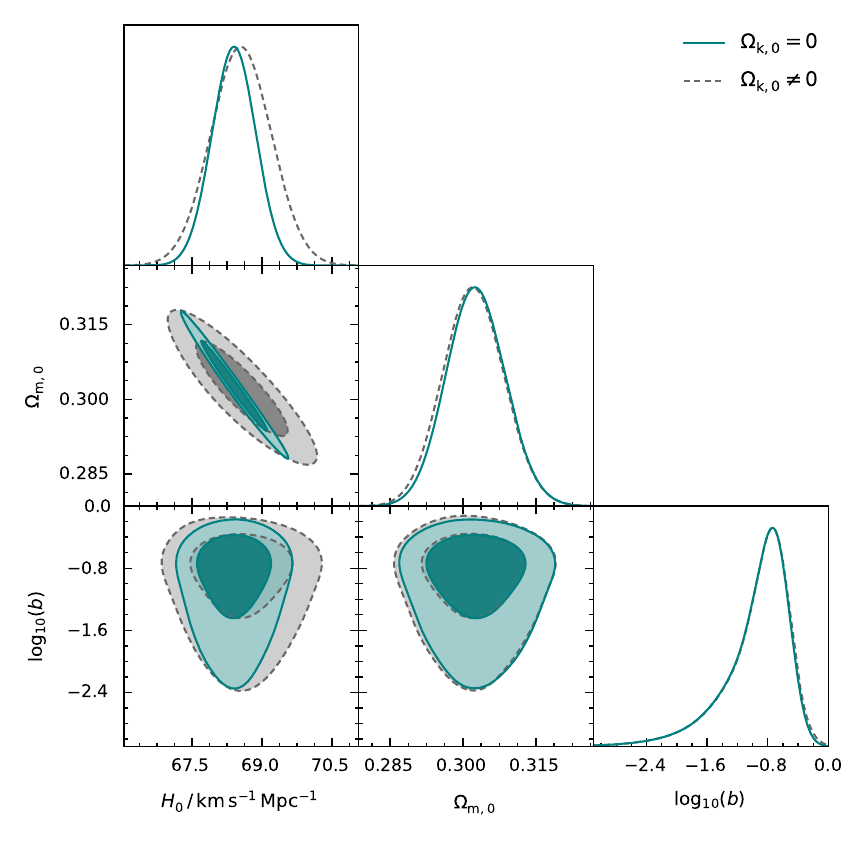}
        \caption{\label{fig:Exp_flat_triangle}{A comparison between the posterior distributions for the main parameters of the spatially flat $(\Omega_{k,0}=0)$ Exponential model and its non-flat $(\Omega_{k,0}\neq0)$ counterpart.}}
    \end{center}
\end{figure}

\begin{figure*}[ht]
    \begin{center}
        \includegraphics[width=0.5\paperwidth]{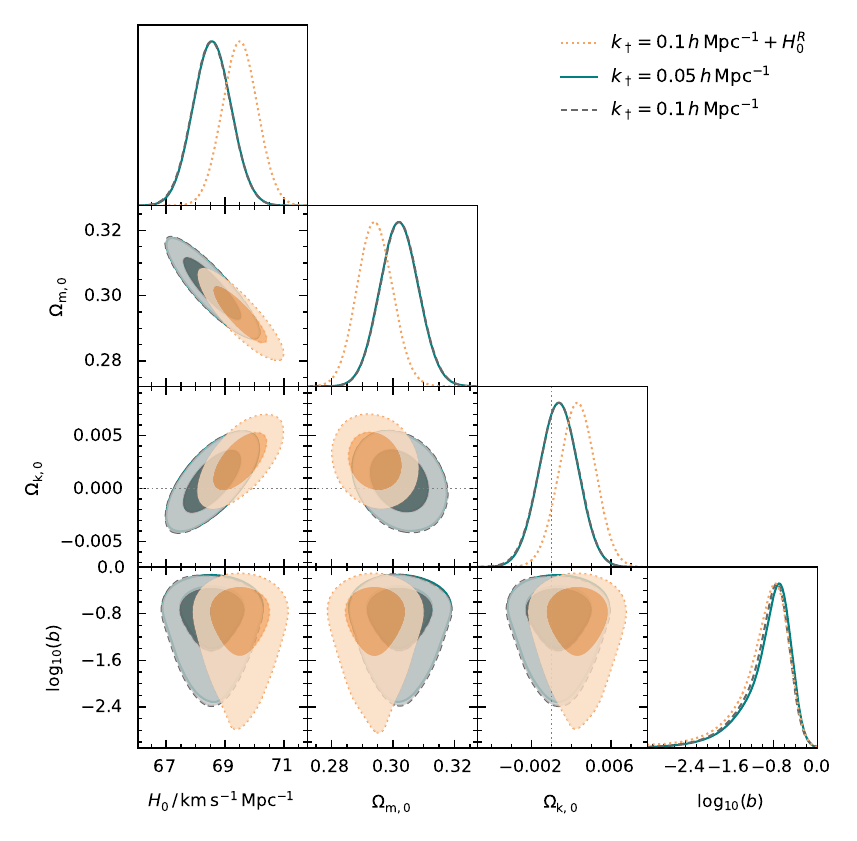}
        \vspace{-3mm}
        \caption{\label{fig:Exp_nonflat_triangle}{Marginalized 2D and 1D posterior distributions for the main parameters of the spatially non-flat Exponential model.}}
    \end{center}
\end{figure*}

\begin{figure}[ht]
    \begin{center}
        \includegraphics[width=0.98\columnwidth]{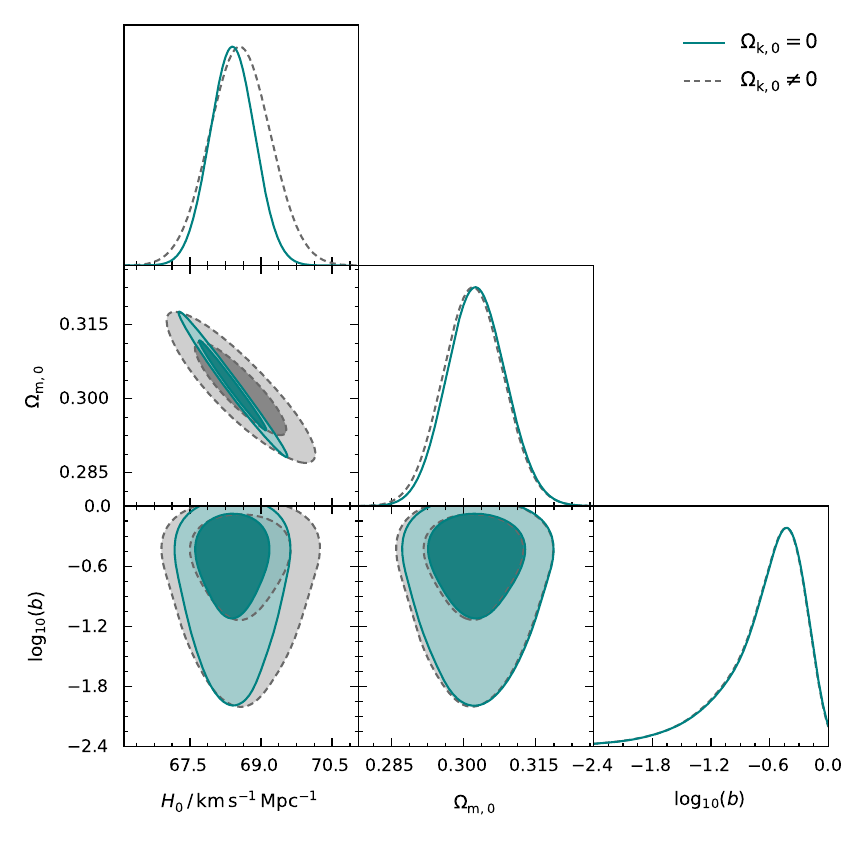}
        \vspace{-3mm}
        \caption{\label{fig:Tsuj_flat_triangle}{A comparison between the posterior distributions for the main parameters of the spatially flat $(\Omega_{k,0}=0)$ Tsujikawa model and its non-flat $(\Omega_{k,0}\neq0)$ counterpart.}}
    \end{center}
\end{figure}

Here, the mean values for the deviation parameter $b$ are significantly larger than the ones obtained in the context of the Hu-Sawicki and Starobinsky models. As may be inferred from Table \ref{T_Exp}, the constraints on $b$ are consistent with zero, i.e.~with the $\Lambda$CDM limit, within a $2\sigma$ confidence interval. Moreover, from Figs.~\ref{fig:Exp_flat_triangle} and \ref{fig:Exp_nonflat_triangle} we observe that the posteriors for $\log_{10}{b}$ are non--Gaussian. Including the $H_0^\mathrm{R}$ likelihood again leads to non--null spatial curvature at a little over $1\sigma$.

\subsubsection{The Tsujikawa model}

\renewcommand{\arraystretch}{1.5}
\begin{table*}[ht]
\center
\caption{\label{T_Tsuji} Mean values and $1\sigma$ confidence intervals for the Tsujikawa model parameters.}
\vspace{1.5mm}
\begin{tabular}{l@{\hskip 1.0cm} A@{\hskip 1.0cm} A@{\hskip 1.0cm} A@{\hskip 1.0cm} A}  
\hline
\hline
Parameter & \multicolumn{2}{c}{Flat~~~~} & \multicolumn{2}{c}{Nonflat~~~~} & \multicolumn{2}{c}{Nonflat ($+H_0^\text{R}$)} & \multicolumn{2}{c}{Nonflat}\\ 
~~ & \multicolumn{2}{c}{$k_\dagger=0.1\,h\,\text{Mpc}^{-1}$} & \multicolumn{2}{c}{$k_\dagger=0.1\,h\,\text{Mpc}^{-1}$} & \multicolumn{2}{c}{$k_\dagger=0.1\,h\,\text{Mpc}^{-1}$} & \multicolumn{2}{c}{$k_\dagger=0.05\,h\,\text{Mpc}^{-1}$} \\ \hline
$H_0$&68.3970&_{-0.4815}^{+0.4811}&68.5620&_{-0.6591}^{+0.6480}&69.4970&_{-0.6144}^{+0.6085}&68.5540&_{-0.6608}^{+0.6552}\\ 
$10^3\Omega_{\mathrm{b},0}^{} h^2$&22.4550&_{-0.1450}^{+0.1448}&22.4220&_{-0.1712}^{+0.1713}&22.4340&_{-0.1734}^{+0.1722}&22.4210&_{-0.1706}^{+0.1706}\\ 
$\Omega_{\mathrm{cdm},0} h^2$&0.1191&_{-0.0010}^{+0.0010}&0.1195&_{-0.0016}^{+0.0015}&0.1197&_{-0.0016}^{+0.0016}&0.1196&_{-0.0016}^{+0.0015}\\ 
$10^3\Omega_{k,0}$&-&-&0.7180&_{-1.9563}^{+1.9853}&2.5897&_{-1.8542}^{+1.8825}&0.7129&_{-1.9624}^{+1.9839}\\ 
$b$&0.3028&_{-0.2950}^{+0.1854} & 0.2999&_{-0.2920}^{+0.1775} & 0.3022&^{+0.1848}_{-0.2965} & 0.3312&_{-0.3235}^{+0.1815}\\ 
$n_s$&0.9697&_{-0.0041}^{+0.0041}&0.9686&_{-0.0051}^{+0.0052}&0.9684&_{-0.0052}^{+0.0052}&0.9686&_{-0.0051}^{+0.0051}\\ 
$\ln{\left(10^{10}A_s\right)}$&3.0443&_{-0.0144}^{+0.0148}&3.0442&_{-0.0145}^{+0.0149}&3.0442&_{-0.0145}^{+0.0150}&3.0442&_{-0.0145}^{+0.0149}\\ \hline
$\Omega_{\Lambda,0}$\,\emph{(derived)}&0.6972&_{-0.0061}^{+0.0063}&0.6971&_{-0.0061}^{+0.0063}&0.7031&_{-0.0057}^{+0.0059}&0.6970&_{-0.0061}^{+0.0063}\\
\hline
\hline
\end{tabular}
\end{table*}


\begin{figure*}[ht]
    \begin{center}
        \includegraphics[width=0.6\paperwidth]{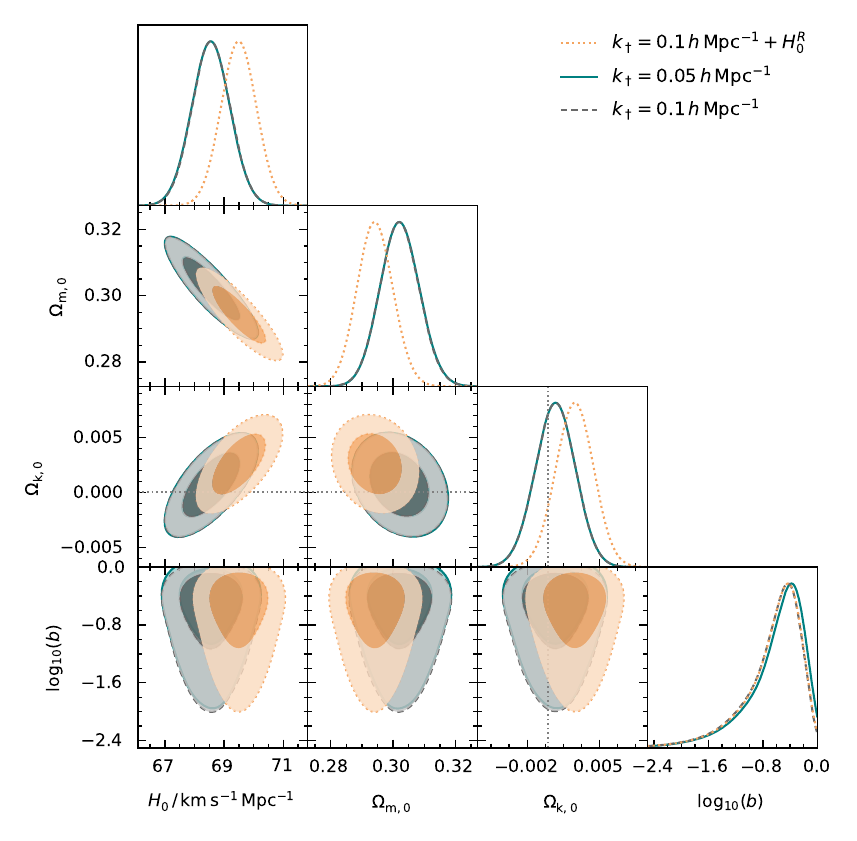}
        \vspace{-3mm}
        \caption{\label{fig:Tsuj_nonflat_triangle}{Marginalized 2D and 1D posterior distributions for the main parameters of the spatially non-flat Tsujikawa model.}}
    \end{center}
\end{figure*}

The derived confidence regions for the main parameters of this model are depicted in Figs.~\ref{fig:Tsuj_flat_triangle} and \ref{fig:Tsuj_nonflat_triangle}, where we represent a selection of pairwise posterior probability functions. We also show the corresponding marginalized 1D distributions. In similar fashion to the Exponential model, the posterior distributions for $\log_{10}{b}$ are characterized by a non--Gaussian profile, and the constraints on $b$ include the $\Lambda$CDM limit, i.e. $b=0$, within a $2\sigma$ confidence interval. 

We report the inferred mean parameter values and the associated uncertainties in Table \ref{T_Tsuji}. As was observed for the other $f(R)$ models, $\Omega_{k,0}$ deviates from zero at a little over $1\sigma$ when we introduce a likelihood for $H_0$. 

\subsubsection{Comparison with $\varLambda$CDM}
Next, we use the same data sets to constrain the parameters of the $\Lambda$CDM model, and consider three cases: 1) $\Omega_{k,0}=0$, 2) freely varying $\Omega_{k,0}$ and 3) freely varying $\Omega_{k,0}$ plus a Gaussian likelihood for $H_0$, with the latter constructed as outlined in Section \ref{sec:observational}. The $f(R)$ models have an extra degree of freedom relative to $\Lambda$CDM (the deviation parameter, $b$). Hence, a robust comparison entails that we use a statistic which also takes this into account, because although more degrees of freedom can mean that the model is better able to approximate the data, this usually comes at the cost of weaker parameter constraints. We therefore employ the Akaike Information Criterion (AIC) \cite{Akaike}:
\begin{equation}
    \text{AIC} = 2p -2\ln{(\cal{L}_\text{max})}~,
\end{equation}
as well as the Bayesian Information Criterion (BIC) \cite{Schwarz}: 
\begin{equation}
    \text{BIC} = p\ln{N}-2\ln{(\cal{L}_\text{max})}~,
\end{equation}
and use them to gauge the performance of the $f(R)$ models in relation to $\Lambda$CDM. In the above equations, $p$ is the amount of free parameters, $N$ the number of observations,\footnote{$N=1105$ in the absence of a likelihood for $H_0$, and $1106$ otherwise.} and $\cal{L}_\text{max}$ denotes the maximum likelihood. When comparing two models, the quantity of interest is the difference between their AIC (BIC) values, as this indicates the level of support for the model with the smaller AIC (BIC): a $|\Delta\text{AIC}|$ that lies in the range from 2 to 4 provides considerable support, while values greater than 10 are effectively conclusive. In the case of the BIC, a difference of magnitude 2 is considered to favour the model having the smaller BIC, and (absolute) differences of 6 or more constitute strong evidence \cite{Nesseris}. 

Our results are presented in Table \ref{BIC_AIC}. The BIC penalizes more heavily for extra parameters, and in fact $\Lambda$CDM comes out on top when the 5 models are compared using this statistic. On the other hand, the AIC values for the Exponential and Tsujikawa models are relatively close to their $\Lambda$CDM counterparts, which implies that these $f(R)$ models perform similarly to $\Lambda$CDM when assessed with the AIC. The Starobinsky model is disfavoured, albeit not strongly. The Hu-Sawicki model, however, appears to be ruled out by both information criteria. It is interesting to note that despite the extra free parameter, the nonflat models get lower AIC and BIC scores than the respective flat ones when no likelihood for $H_0$ has been included in the analysis. The Starobinsky model is the only one that does not follow this trend. 

The affinity of the Exponential and Tsujikawa models to $\Lambda$CDM concurs with the fact that for the greater majority of the cosmological evolution, they are essentially identical to the standard model. It is only at very late times that deviations set in, and when they do, they provide a neat mechanism for dark energy that is still painfully lacking in $\Lambda$CDM. These two models augment the benefits of $\Lambda$CDM with a well-motivated theoretical basis for the observed acceleration of the Universe, and in this sense are especially appealing. 

\renewcommand{\arraystretch}{1.5}
\begin{table*}[ht]
\center
\caption{\label{BIC_AIC} The AIC and BIC statistics for the Hu-Sawicki, Starobinsky, Exponential, Tsujikawa and $\Lambda$CDM models. $\Delta$AIC and $\Delta$BIC are calculated by using the AIC and BIC values for $\Lambda$CDM as baseline. The constraints on the parameters of the standard model are independent of $k_\dagger$, so in this case we do not distinguish between the scenarios with $\Omega_{k,0}\neq 0,~k_\dagger=0.1\,h\,\text{Mpc}^{-1}$ and $\Omega_{k,0}\neq 0,~k_\dagger=0.05\,h\,\text{Mpc}^{-1}$.}
\vspace{1.5mm}
\begin{tabular}{K{2cm} K{2cm} K{3cm} K{3cm} K{3cm} K{3cm}}
\hline
\hline
Model & Statistic & Flat & Nonflat & {Nonflat ($+H_0^\text{R}$)} & Nonflat\\ 
~~ & ~~ & {$k_\dagger=0.1\,h\,\text{Mpc}^{-1}$} & {$k_\dagger=0.1\,h\,\text{Mpc}^{-1}$} & {$k_\dagger=0.1\,h\,\text{Mpc}^{-1}$} & {$k_\dagger=0.05\,h\,\text{Mpc}^{-1}$} \\ \hline
Hu-Sawicki & AIC & 1089.02 & 1082.37 & 1106.86 & 1085.78\\
~ & $\Delta$AIC & 13.48 & 4.36 & 16.21 & 7.77\\
~ & BIC & 1124.08 & 1122.43 & 1146.92 & 1125.84\\
~ & $\Delta$BIC & 18.50 & 9.37 & 21.21 & 12.78\\
\hline
Starobinsky & AIC & 1077.55 & 1081.38 & 1093.36 & 1080.45\\
~ & $\Delta$AIC & 2.01 & 3.37 & 2.71 & 2.44\\
~ & BIC & 1112.60 & 1121.44 & 1133.43 & 1120.51\\
~ & $\Delta$BIC & 7.02 & 8.38 & 7.72 & 7.45\\
\hline
Exponential & AIC & 1077.13 & 1079.01 & 1092.32 & 1079.51\\
~ & $\Delta$AIC & 1.59 & 1.00 & 1.67 & 1.50\\
~ & BIC & 1112.18 & 1119.07 & 1132.39 & 1119.57\\
~ & $\Delta$BIC & 6.60 & 6.01 & 6.68 & 6.51\\
\hline
Tsujikawa & AIC & 1077.59 & 1079.66 & 1092.06 & 1079.51\\
~ & $\Delta$AIC & 2.05 & 1.65 & 1.41 & 1.50\\
~ & BIC & 1112.64 & 1119.72 & 1132.13 & 1119.57\\
~ & $\Delta$BIC & 7.06 & 6.66 & 6.42 & 6.51\\
\hline
$\Lambda$CDM & AIC & 1075.54 & 1078.01 & 1090.65 & 1078.01\\
~ & BIC & 1105.58 & 1113.06 & 1125.71 & 1113.06\\
\hline
\hline
\end{tabular}
\end{table*}

\section{Conclusion}
\label{sec:conclusion}

\begin{figure}
\begin{center}
\includegraphics[width=0.95\columnwidth]{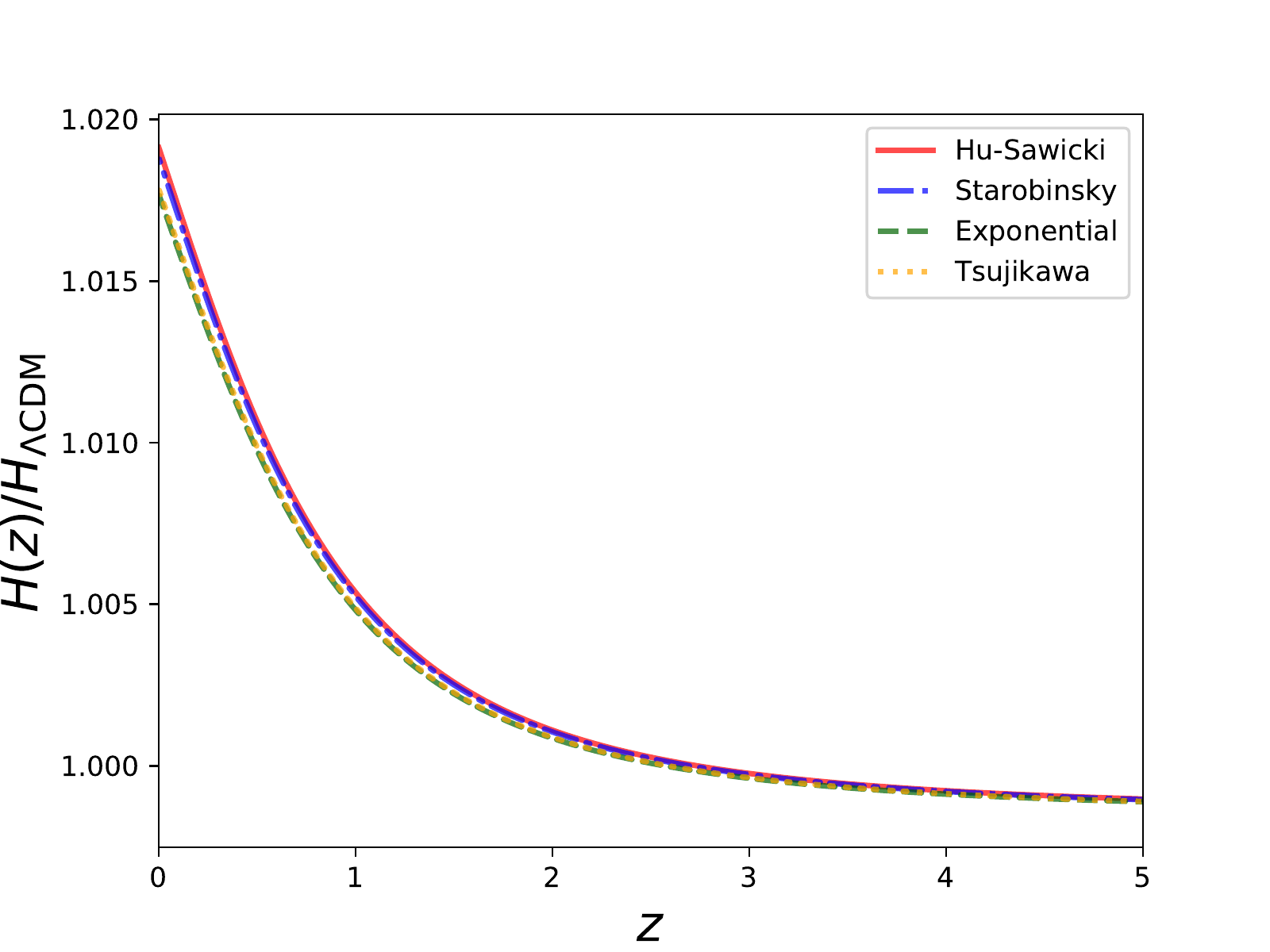}
\vspace{-2mm}
\caption{\label{hubble_ratio_nonflat_case}{The variation of $H(z)/H_{\Lambda\text{CDM}}$ with redshift at late times. $H(z)$ was calculated using the mean values from the third column of Tables \ref{T_HS}--\ref{T_Tsuji}, and thus describes the expansion of a universe with $\Omega_{k,0}\neq 0$. $H_{\Lambda\text{CDM}}$ corresponds to a $\Lambda$CDM cosmology in which the density parameters and $H_0$ take the \emph{Planck} TT,TE,EE+lowE+lensing mean values, and $\Omega_{k,0}=0$. Therefore, the ratio $H(z)/H_{\Lambda\text{CDM}}$ does not go to unity at high redshifts.}}
\end{center}
\vspace{-2mm}
\end{figure}

\begin{figure}
\begin{center}
\includegraphics[width=0.93\columnwidth]{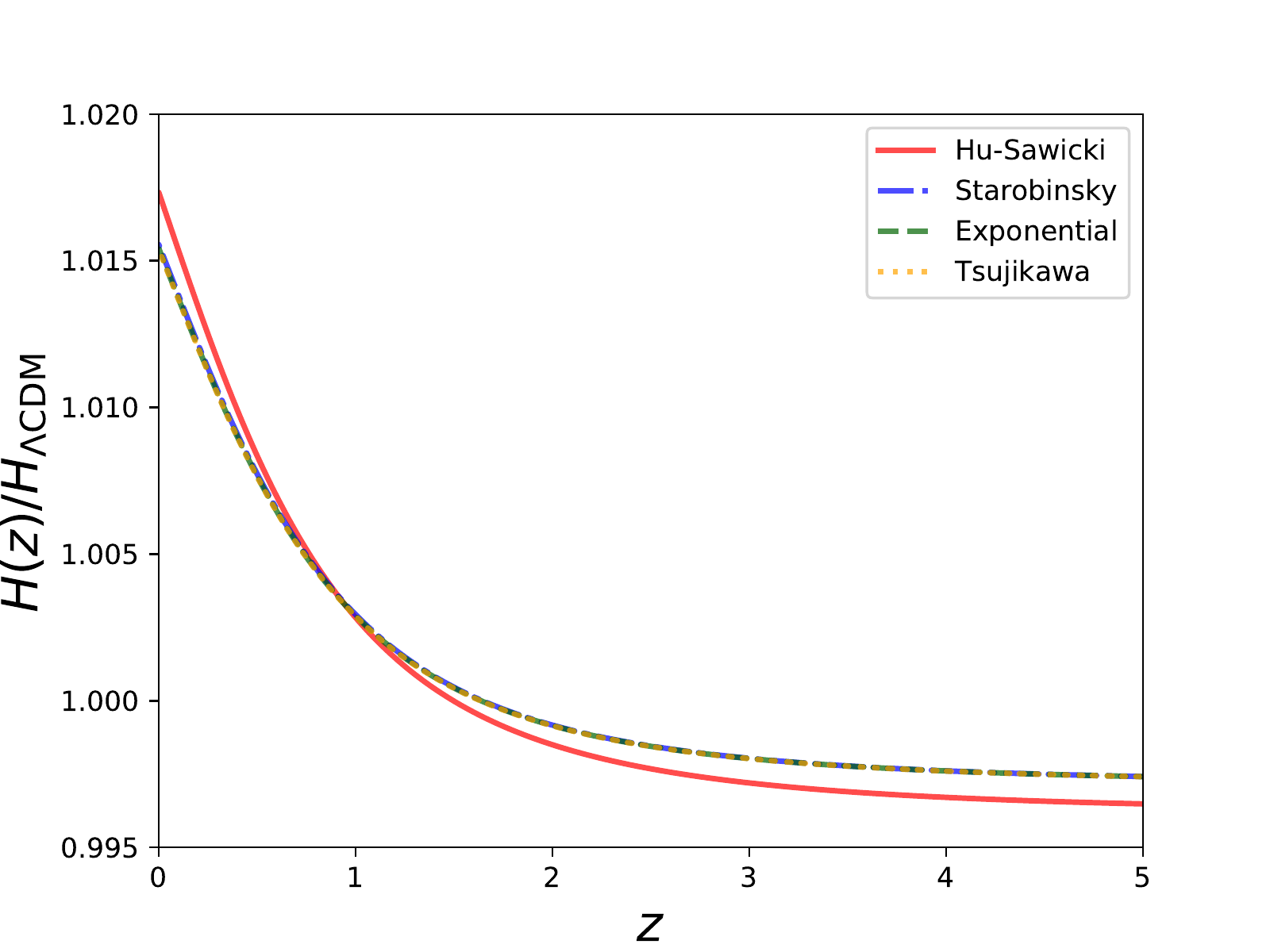}
\caption{\label{hubble_ratio_flat_case}{Same as for Fig.~\ref{hubble_ratio_nonflat_case}, except that in this case, $H(z)$ is the Hubble parameter for a spatially flat universe, and was obtained using the mean values from the second column of Tables \ref{T_HS}--\ref{T_Tsuji}.}}
\end{center}
\end{figure}

The work presented here focuses on four $f(R)$ models and places constraints on their parameters by means of data from SNeIa, the CMB, BAOs, cosmic chronometers and RSDs. 

The action of $f(R)$ gravity is constructed from that of GR by generalizing the Ricci scalar $R$ to a function thereof. This results in a model that can be seen as a natural extension of GR, and that can furthermore explain the current accelerated expansion of the Universe without any need for dark energy. We study $f(R)$ gravity in the context of the Hu-Sawicki \cite{Hu2007}, Starobinsky \cite{Starobinsky}, Exponential \cite{Cognola} and Tsujikawa \cite{STsujikawa2008} models. A common feature of these four models is the fact that the respective $f(R)$ functions can all be expressed in the form $f(R)=R-2\Lambda X$, where $\Lambda$ is the cosmological constant and $X$ represents a quantity that goes to unity at high redshifts. Thus, any differences from $\Lambda$CDM emerge at late times. 

The main aim of our work is to investigate in what ways, if any, the behavior of the models changes when the assumption of spatial flatness is relaxed. To this end, we use MCMC techniques to sample the parameter space of both cosmological and model-specific parameters, initially putting $\Omega_{k,0}=0$, then allowing it to vary. As expected, constraints on cosmological parameters tend to be tighter when $\Omega_{k,0}$ is set to a fixed value. To further probe the role of curvature, we plot the ratio $H(z)/H_{\Lambda\text{CDM}}$ for each of the four models over the redshift range $[0,5]$. $H(z)$ is evaluated by using the mean parameter values listed in Tables \ref{T_HS}--\ref{T_Tsuji}, whereas $H_{\Lambda\text{CDM}}$ corresponds to a flat $\Lambda$CDM cosmology whose parameters are assigned the \emph{Planck} values from Ref.~\cite{Planck2018VI}. Fig.~\ref{hubble_ratio_nonflat_case} shows that in the non-flat case (no $H_0$ likelihood), the departure from $\Lambda$CDM is comparable across the four models; any variations at low redshifts mainly arise from the fact that the models have a different Hubble constant. We have checked that this also holds if $k_\dagger=0.05\,h\,\text{Mpc}^{-1}$. Similar behavior is noted when we adopt the parameter values from the fourth column of Tables \ref{T_HS}--\ref{T_Tsuji}, i.e.~those obtained with an additional likelihood -- a normal distribution constructed from the local measurement of $H_0$. Now, however, the maximum value of $H(z)/H_{\Lambda\text{CDM}}$ increases, reflecting the higher averages we get for $H_0$ in the presence of said likelihood. In the flat scenario, the curve for the Hu-Sawicki model stands out from the rest at redshifts $z>1$ (Fig.~\ref{hubble_ratio_flat_case}). This is due to a larger deviation from $\Lambda$CDM, which may in turn be attributed to a smaller mean value of $\Omega_{\text{cdm},0}h^2$ (a quantity equal to $0.1188$, compared to $0.1191$ for the other models).\footnote{We point out that the separation between the Hu-Sawicki and the three other curves in Fig.~\ref{hubble_ratio_flat_case} (for redshifts $z>1$) is, in fact, not much bigger than that between the upper- and lower- most curves in some of the other cases we consider. The crucial difference is that here we do not get a gradual change from model to model.}  

On the whole, our results are consistent with spatial flatness. We note, nonetheless, that the constraints obtained upon adding $H_0^\text{R}$ to the data set favor an open universe at a little over $1\sigma$. This is in line with the fact that $H_0$ is correlated with $\Omega_{k,0}$, so a higher mean value for the Hubble constant translates into a larger $\Omega_{k,0}$. It would seem prudent, therefore, not to exclude spatial curvature before the nature of the Hubble tension has been clarified. Indeed, model-independent estimates of $\Omega_{k,0}$, obtained in Ref.~\cite{Ruan} using cosmic chronometer data and a Gaussian Process reconstruction of the HII galaxy Hubble diagram, show that while the \emph{Planck} value for the Hubble constant lends support to a flat universe, the local measurement of $H_0$ rules it out at around $3\sigma$.
~~~~~~~~~~~~~~~~~~\\

\begin{acknowledgments}
J.~M.~would like to acknowledge funding support from Cosmology@MALTA (University of Malta) and KASI. The research work of C.~F.~disclosed in this publication was partially funded by the Endeavour Scholarship Scheme (Malta). Scholarships are part-financed by the European Union -- European Social Fund (ESF) -- Operational Programme II -- Cohesion Policy 2014--2020: \emph{``Investing in human capital to create more opportunities and promote the well-being of society''}. Numerical computations were carried out on the Sciama High Performance Compute (HPC) cluster, which is supported by the ICG, SEPNet and the University of Portsmouth.
\end{acknowledgments}

\bibliography{biblio_library}

\end{document}